\documentclass[twocolumn]{aastex631}
\usepackage{multirow}
\usepackage{makecell}
\usepackage[table]{xcolor}

\makeatletter
\renewenvironment{acknowledgments}{%
  \section*{ACKNOWLEDGMENTS}
}{%
}
\makeatother

\usepackage{xspace}
\usepackage{colortbl}
\usepackage{xcolor}
\usepackage{subfigure}
\usepackage{chemformula}
\usepackage{booktabs}
\usepackage{hyperref}
\usepackage{amsmath}
\usepackage{tabularx}
\usepackage{array}
\usepackage{float}

\DeclareSymbolFont{UPM}{U}{eur}{m}{n}
\DeclareMathSymbol{\umu}{0}{UPM}{"16}
\newcommand\micro{$\umu$}
\newcommand\microns{\micro m\xspace}

\def\rfast{\texttt{\footnotesize{rfast}}\xspace}
\def\pyedith{\texttt{\footnotesize{pyEDITH}}\xspace}
\def\emcee{\texttt{\footnotesize{emcee}}\xspace}

\submitjournal{The Astrophysical Journal}
\accepted{April 27, 2026}

\begin{document}

\title{The effect of spectral resolution on biosignature detection via reflected light observations of the Earth through time}

\correspondingauthor{Samantha Gilbert-Janizek}
\email{samroseg@uw.edu}

\author[0009-0004-8402-9608]{Samantha Gilbert-Janizek}
\affiliation{Department of Astronomy and Astrobiology Program, University of Washington, Box 351580, Seattle, Washington 98195}

\author[0000-0002-0746-1980]{Jacob Lustig-Yaeger}
\affiliation{Johns Hopkins Applied Physics Laboratory, Laurel, Maryland 20723, USA}
\affiliation{NASA NExSS Virtual Planetary Laboratory, Box 351580, University of Washington, Seattle, Washington 98195, USA}

\author[0000-0001-6878-4866]{Joshua Krissansen-Totton}
\affiliation{Department of Earth and Space Sciences and Astrobiology Program, University of Washington, Seattle, WA 98195, USA.}
\affiliation{NASA NExSS Virtual Planetary Laboratory, Box 351580, University of Washington, Seattle, Washington 98195, USA}

\begin{abstract}
NASA's Habitable Worlds Observatory (HWO) will search for biosignatures on Earth-like exoplanets using reflected light spectroscopy. A critical instrument design parameter is resolving power, which must balance biosignature detectability against exposure time and detector noise constraints. We assess the resolving power needed to detect and characterize key biosignature gases and habitability indicators including \ch{O2}, \ch{O3}, \ch{H2O}, \ch{CH4}, \ch{CO2} and \ch{CO} across atmospheres representing the Archean, Proterozoic, and Phanerozoic Earth. We combine analytical detectability calculations spanning spectral resolutions ($\lambda/\Delta{\lambda}$) $R=20$--$5000$ with atmospheric retrievals using the \rfast radiative transfer model and \pyedith exposure time calculator for realistic wavelength-dependent noise modeling. In the visible ($0.4$--$1.0$ \microns), the nominal resolution $R_{\mathrm{Vis}}=140$ is sufficient for detecting \ch{O2} in Phanerozoic-like atmospheres. Higher resolutions could theoretically reduce exposure times for low-\ch{O2} Proterozoic atmospheres, but require $>$$10{\times}$ reductions in dark current and could increase \ch{H2O} detection exposure times by ${\sim}2{\times}$, penalizing the foundational habitability constraint that anchors downstream biosignature searches. The most efficient path for low-\ch{O2} atmospheres may instead be indirect inference via \ch{O3}, whose Hartley-Huggins bands are detectable at $R_{\mathrm{UV}}{\sim}7$. In the near-IR ($1.0$--$1.7$ \microns), $R_{\mathrm{NIR}}\geq40$ is necessary to avoid a degeneracy between \ch{CO2} and \ch{CO} that could produce false positive detections of abundant \ch{CO}. The nominal $R_{\mathrm{NIR}}=70$ is sufficient for characterizing all Earth-through-time cases. These results support HWO's current baseline resolution choices and provide actionable guidance for finalizing spectrometer requirements while maintaining technological feasibility for the search for life on exoplanets.
\end{abstract}


\section{Introduction} \label{sec:intro}

NASA's next flagship mission, the Habitable Worlds Observatory (HWO), will enable the search for life on Earth-like exoplanets. HWO is planned to be a $\sim$6-meter space-based ultraviolet (UV), optical, and near-infrared (NIR) direct imaging telescope that will detect and characterize the atmospheres of nearby Earth-like exoplanets around Sun-like (FGK) stars \citep{feinberg2026habitable}. A major priority of HWO will be to use reflected light spectroscopy to identify potentially habitable and inhabited planets \citep{national2021pathways}. Potentially inhabited planets may be identified by the detection of surface and atmospheric biosignature gases, which may include biological pigments or the metabolic products of life, respectively \citep{des2002remote, seager2005vegetation, meadows2018exoplanet, schwieterman2018exoplanet, schwieterman2024overview}.

Though living planets and their biosignatures may be incredibly diverse, life's ${\sim}4$ Gyr history on Earth acts as an informative baseline for mission design. Prior to the emergence of widespread photosynthetic life ${\sim}2.4$ Ga, the anoxic Archean Earth atmosphere included abundant methane largely produced by life \citep{catling2001biogenic, kasting2005methane, thompson2022case, ebadirad2025archean}. While atmospheric oxygen levels grew following the Great Oxidation Event ${\sim}2.4$ Ga, the Proterozoic Earth atmosphere likely had \ch{O2} and \ch{O3} abundances orders of magnitude smaller than today \citep{kump2008rise, lyons2014rise, planavsky2014low, bellefroid2018constraints}. The Phanerozoic Earth atmosphere, with 20\% \ch{O2}, strong UV shielding due to stratospheric \ch{O3}, and trace abundances of \ch{CH4} {is practically identical to} the animal and plant-dominated planet we know today {, more commonly referred to as the Modern Earth in previous work \citep[e.g.,][]{gomez2023search, young2024retrievals}}. Atmospheric water vapor and \ch{CO2}, two possible indicators of habitability, have been present throughout Earth's history \citep{catlingdavid2018exoplanet}. Based on Earth's biogeochemical history, important molecules for detecting life on exoplanets thus include \ch{O2}, \ch{O3}, \ch{H2O}, \ch{CH4}, and \ch{CO2}. However, the detection of these molecules alone does not constitute a confident life detection due to the possibility of false positive atmospheres \citep{domagal2014abiotic, meadows2018exoplanet, krissansen2021oxygen, thompson2022case}.  

False positive atmospheres mimic life's presence via abiotic generation of biosignature gases, and the effort to identify possible molecular discriminants to rule out such scenarios is ongoing. For methanogenic life, recent work evaluating possible detections of \ch{CH4} on Mars argued that abundant \ch{CO} would imply an abiotic origin \citep{zahnle2011there}, since CO is a more energetically advantageous source of carbon than \ch{CO2} \citep{ragsdale2004life}. Building on this argument, recent work proposed that the presence of abundant \ch{CO} alongside \ch{CH4} and \ch{H2O} is one possible ``antiobiosignature'' for anoxygenic atmospheres \citep{wogan2020chemical, krissansen2022understanding}. Furthermore, in reducing biospheres, \citet{schwieterman2019rethinking} showed that biomass burning may only plausibly produce \ch{CO} abundances as high as 10 ppm on planets around FGK stars. \ch{CO} may also be relevant to the identification of oxygenic biospheres. \citet{hu2020o2} showed that photolysis in habitable, \ch{CO2}-rich atmospheres could lead to the abiotic accumulation of both \ch{CO} and \ch{O2}, potentially producing a false positive biosignature scenario where oxygen is present without life. However, in follow-up work \citet{ranjan2023importance} demonstrated that this result was likely a model artifact related to the upper atmosphere boundary condition, finding that \ch{O2} would remain at trace abundances rather than accumulating to modern-Earth-like levels, thereby preserving \ch{CO}'s utility as an antibiosignature for Earth-through-time analogs.





To ensure that HWO is capable of finding life on exoplanets, the community is considering how different observatory architectures and baseline instrument parameters affect our ability to detect and constrain these key gases. To observe and characterize comparatively dim exoplanet atmospheres around FGK stars, HWO will be equipped with a coronagraph and reflected light spectroscopy capable of UV/optical/NIR observations. In addition to spectroscopic wavelength range and band-pass signal-to-noise ratios (SNRs) \citep{krissansen2025wavelength}, we must determine the spectral resolving power required to characterize the atmospheres of Earth-like exoplanets and ultimately identify habitable and/or living planets \citep{luvoir2019luvoir, gaudi2020habitable}. 

Previous work argued that HWO's spectral resolution ($\lambda/\Delta\lambda$) in the visible (0.4 -- 1.0 \microns) should be set by the detection of the narrow-band \ch{O2} feature at 0.76 \microns \citep{des2002remote, brandt2014prospects, feng2018characterizing, luvoir2019luvoir, gaudi2020habitable, damiano2022reflected}, while the resolution in the near-IR (1.0 -- 1.7 \microns) should be set by the detection of \ch{CH4} and \ch{CO2} features \citep{des2002remote, luvoir2019luvoir, gaudi2020habitable, damiano2022reflected}. These recommendations have emerged from studies using a variety of analysis techniques, including measurements of band-pass absorption features \citep{des2002remote}, gas-specific detectability calculations \citep{brandt2014prospects}, and finally atmospheric retrieval simulations \citep{feng2018characterizing, damiano2022reflected}. 


The exploration of observatory trade space has benefited enormously from the application of spectral retrieval models because they help assess our ability to meaningfully constrain gas abundances \citep{damiano2021reflected, damiano2022reflected, damiano2023reflected, latouf2023bayesian, latouf2023bayesian2, gilbert2024retrieved, tokadjian2024detectability, latouf2025barbie, young2025modern, krissansen2025wavelength, hagee26}. Abundance constraints provide unique information about the character of planetary atmospheres that detections alone cannot. Recently, \citet{feng2018characterizing} and \citet{damiano2022reflected} used retrieval models to assess how two different spectral resolutions in the visible impact our ability to characterize Earth-like atmospheres. Both studies explored how $R=70$ and $R=140$ in the visible impact our ability to detect and constrain \ch{O2}, \ch{O3}, and \ch{H2O}. Additionally, \citet{damiano2021reflected} showed that including $R=40$ observations in the near-IR wavelengths provides important contextual information about atmospheric \ch{CO2}. The LUVOIR final report recommended $R_{\mathrm{NIR}}=70$ for \ch{CO2} detection with an integral field spectrograph and $R_{\mathrm{NIR}}=200$ for single-point spectrographs \citep{luvoir2019luvoir}. During the preparation of this manuscript, \citet{ruffio2026characterizing} quantified the effects of correlated noise due to speckles, recommending higher resolutions ($R{\sim}1000$) may increase sensitivity in the optical to \ch{O2} and \ch{H2O} under the plausible assumption that the speckle correlation length is on the order of (or smaller than) the width of the spectral features. However, no retrieval study to-date has fully and systematically explored the effect of spectral resolution in the visible and the near-IR on our ability to detect and constrain biosignatures for Earth-through-time atmospheres with HWO.

In this study, we combine forward modeling with analytical detectability calculations and Bayesian retrieval simulations. We generate synthetic HWO observations across a range of spectral resolutions ($R = 20$--$5000$ in the visible; $R = 20$--$1000$ in the near-infrared) using realistic wavelength-dependent noise models. Our analytical approach efficiently explores the full resolution trade space by calculating minimum exposure times required for confident biosignature detections via Bayesian Information Criterion thresholds. We then validate and extend these findings through full atmospheric retrievals on select cases of particular interest, which reveal how resolution affects not just detectability but our ability to meaningfully constrain gas abundances, including critical upper limits on potential antibiosignatures. This dual approach allows us to answer the following time-critical questions: (1) Can higher visible spectral resolution compensate for the challenges of detecting low \ch{O2} in Proterozoic atmospheres, either by reducing required exposure times or by providing an alternative to UV coronagraphy? (2) Can we reduce near-IR spectral resolution without compromising atmospheric characterization? (3) How do these trade-offs change across different stages of Earth's atmospheric evolution?

In \autoref{sec:methods}, we describe our atmospheric test cases spanning Earth's history, the \rfast radiative transfer model used to generate synthetic spectra, the \pyedith exposure time calculator for realistic noise modeling, our analytical framework for calculating detectability thresholds, and our retrieval methodology. In \autoref{sec:results}, we present analytical calculations of exposure time versus resolution for all Earth-through-time cases, followed by detailed retrieval simulations for the Proterozoic and Phanerozoic (visible wavelengths) and Archean and Phanerozoic (near-infrared wavelengths). We discuss our findings in the context of previous studies and HWO mission design in \autoref{sec:discussion}, and conclude with our spectral resolution recommendations in \autoref{sec:conclusion}.

\section{Methods}
\label{sec:methods}

\subsection{Atmospheric Cases}

We consider the same Earth-through-time cases as described by \citet{krissansen2025wavelength}, which are summarized in \autoref{tab:atms}. Accordingly, our study complements their investigation of the short-wavelength and long-wave cutoffs needed to accurately characterize the Earth through time. Mixing ratios are chosen based on literature estimates appropriate for each Earth epoch \citep[See][Sec. 2.2 for a thorough description]{krissansen2025wavelength}. Generally, our atmospheres are characterized by secularly decreasing \ch{CO2} with \ch{CO}$<$\ch{CH4} due to biogenic CO consumption. Our Archean atmospheres have higher \ch{CH4} abundances (0.5--0.05\%) due to the predominance of methanogenic life. Our Proterozoic atmospheres have 0.2\% \ch{O2} abundances (1\% PAL) due to the emergence of oxygenic life, which rapidly oxidized the atmosphere. The oxidized atmosphere readily destroys \ch{CH4}, leading to lower abundances in the Proterozoic and Phanerozoic eons. For the Proterozoic and Archean epochs, we explore both ``high'' and ``low'' \ch{CH4} scenarios. All atmospheres are assumed to be 1 bar. 

\begin{table*}[htbp]
\setlength{\tabcolsep}{16pt}
\centering
\caption{Atmospheric composition and properties of Earth through different geological eons. For gases, abundances are reported as volume mixing ratios. For all cases, we simulate planets with 1.0$R_{\oplus}$, 1.0$M_{\oplus}$, and gray surface albedos of 0.1. Additionally, all atmospheres include 50\% cloud coverage, where cloud decks are assumed to be $1\times10^4$ Pa thick with a cloud-top pressure of $6\times10^4$ Pa and extinction opacity of $\tau_c = 1$.}
\begin{tabular}{lccccc}
\toprule
 & Phanerozoic & Proterozoic & Proterozoic & Archean & Archean \\
 &  & (low \ch{CH4}) & (high \ch{CH4}) & (low \ch{CH4}) & (high \ch{CH4}) \\
\midrule
\ch{N2} & 79.6\% & 79\% & 98.8\% & 80\% & 94.5\% \\
\ch{O2} & 20\% & 0.2\% & 0.2\% & None & None \\
\ch{O3} & $7\times10^{-7}$ & $7\times10^{-8}$ & $7\times10^{-8}$ & None & None \\
\ch{CO2} & 0.1\% & 1\% & 1\% & 20\% & 5\% \\
CO & 0.1 ppm & 0.1 ppm & 0.1 ppm & 0.005\% & 0.05\% \\
\ch{CH4} & 1 ppm & 5 ppm & 100 ppm & 0.05\% & 0.5\% \\
\ch{H2O} & 0.3\% & 0.3\% & 0.3\% & 0.3\% & 0.3\% \\
Surface Pressure (bar) & 1 & 1 & 1 & 1 & 1 \\
\bottomrule
\label{tab:atms}
\end{tabular}
\end{table*}

\subsection{\rfast Forward Model}

To calculate all reflected light spectra in this study, we use the \rfast radiative transfer suite, a 1-D exoplanet atmosphere forward model from \citet{robinson2023exploring}. To calculate molecular opacities based on the the HITRAN2020 database \citep{gordon2022hitran2020}, \rfast uses the Line-By-Line ABsorption Coefficients tool \citep[LBLABC,][]{meadows1996ground}. The model also includes a blended  ice/liquid water cloud model \citep{robinson2023exploring}, which we apply by assuming 50\% cloud coverage for all atmospheric cases. All spectra are initially calculated at a resolution of 5000 in the visible and 1000 in the near-IR wavelength ranges, and subsequently degraded to the resolution under investigation via a Gaussian convolution kernel. Based on the conclusions of \citet{krissansen2025wavelength}, we adopt a long-wavelength cutoff in the near-IR of 1.7 \microns. In contrast to \citet{krissansen2025wavelength}, we adopt a short-wavelength cutoff in the visible of 0.4 \microns, due to the current UV modeling limitations of our noise model (see below).

\subsection{\pyedith Exposure Time Calculator}

We use the Python-based Exposure Direct Imaging Timer for HWO (\pyedith; \citealp{Alei2025}; \citealp{aleicurrie}) to perform noise calculations that are applicable to the current working design of the HWO coronagraph instrument. The \pyedith exposure time calculator (ETC) has heritage from the ETC of \citet{Stark2014}, but represents a new standalone tool for developing and refining exoplanet detection and characterization science capabilities for HWO. \pyedith features the ability to load EAC YAML files released by the HWO project office that specify the observatory and instrument parameters for each Exploratory Analytic Case (EAC)\footnote{{https://github.com/HWO-GOMAP-Working-Groups/Sci-Eng-Interface}}. For more details on \pyedith, please refer to \citet{Alei2025} and the online code documentation\footnote{{https://pyedith.readthedocs.io/en/latest/index.html}}. 

In this work, we use the EAC1 observatory preset, which is provided with the initial release of \pyedith, and we use the IFS observing mode to simulate planetary reflected light spectroscopy using the coronagraph. We provide a list of relevant parameters in \autoref{tab:pyedith} that are used throughout this study unless otherwise noted. As in recent work, we assume that the nominal dark current is $3 \times 10^{-5}$ $e^-$/pixel/second \citep{stark2025cross}, however, we explore a broader range of values in \autoref{sec:results} to investigate how detector noise trades with spectral resolution. We omit UV noise modeling from our investigations due to the limited capability of \pyedith to accurately simulate UV observations at the time of writing. 
Finally, we run \pyedith noise calculations for each of the follow spectral resolving powers: $R_{\mathrm{Vis}}$=30, 50, 70, 100, 140, 200, 500, 1000, 2000, 3000, 4000, and 5000, and $R_{\mathrm{NIR}}$=20, 40, 50, 60, 70, 80, 100, 140, 200, 500, and 1000. These fiducial calculations are for an Earth at 10 pc observed with HWO for 10 hours. We rescale the exposure time analytically throughout this work to investigate the relationship between spectral resolution and the minimum required exposure time to detect biosignature gases and habitability indicators in specific bands of interest. 

To test the validity of re-scaling a single noise model output, we generated noise instances at the nominal resolution for exposure times $t_{\mathrm{exp}} = 5$, $100$, and $1000$ hours, and compared the resulting SNR of our re-scaled exposure times to the truth. We found that our analytical relationship accrues errors in the absolute exposure time no larger than 20\% relative to the regenerated noise instances. The largest exposure time errors are for the shortest exposure times where fixed overhead due to slew, settling, and digging the dark hole take a significant fraction of the total observation. Cases with longer than our nominal 10 hour exposure time minimize the effect of overheads and approach our analytic calculations. Therefore, while these effects are minimal, we caution that our reported absolute exposure times are less accurate than the relative differences between our various cases. Given that the HWO architecture is not yet set, we argue that relative exposure time differences are the important figure of merit, and as such we focus our interpretation of results on them in \autoref{sec:discussion}.  

\begin{table*}
    \centering
    \begin{tabular}{lll}
       \hline 
       Parameter & Value & Meaning \\
       \hline
       \hline
       \multicolumn{3}{c}{Observation} \\
       \hline
       \\[-8pt]
       \texttt{full wavelength range} [{\microns}] & [0.4 -- 1.7] &\parbox{10cm}{Full wavelength range of the observations} \\
       \texttt{visible wavelength range} [{\microns}] & [0.4 -- 1.0] &\parbox{10cm}{Range of wavelengths in the visible} \\
          \texttt{near-IR wavelength range} [{\microns}] & [1.0 -- 1.7] &\parbox{10cm}{Range of wavelengths in the near-infrared} \\
       \texttt{exposure\_time} [hours] & 10 &\parbox{10cm}{Exposure time for the fiducial observation} \\ 
       \texttt{CRb\_multiplier} & 2 & \parbox{10cm}{Factor $\alpha$ for PSF subtraction method (2 for Angular Differential Imaging)} \\
       \texttt{distance} [pc] & 10 & \parbox{10cm}{Distance $d$ of the system} \\
       \texttt{stellar\_radius} [$R_{\odot}$] & 1 & \parbox{10cm}{Stellar radius} \\
       \texttt{nzodis} [zodi] & 3 & \parbox{10cm}{Exozodi multiplier} \\
       \texttt{ra} & $236.01^{\circ}$ & \parbox{10cm}{Right ascension} \\
       \texttt{dec} & $2.52^{\circ}$ & \parbox{10cm}{Declination} \\
       \texttt{angular\_separation} [arcsec] & 0.1 & \parbox{10cm}{Separation (at $d$)} \\
       \hline
       \multicolumn{3}{c}{Telescope} \\
       \hline
       \\[-8pt]
       \texttt{diameter} [m] & 7.226 & \parbox{10cm}{Circumscribed diameter $D_{\mathrm{circ}}$ of telescope aperture} \\
       \texttt{unobscured\_area} [\%] & 100.0 & \parbox{10cm}{Percentage of unobscured collecting area} \\
       \texttt{toverhead\_multi} & 1.100 & \parbox{10cm}{Multiplicative overhead time $\tau_{\mathrm{multi}}$ (refining the dark hole)} \\
       \texttt{toverhead\_fixed} [s] & 8250 & \parbox{10cm}{Static overhead time $\tau_{\mathrm{static}}$ (slew, settling, digging the dark hole)} \\
       \texttt{telescope\_optical\_throughput} & 0.822 & \parbox{10cm}{Throughput of the telescope optics} \\
       \texttt{temperature} [K] & 290 & \parbox{10cm}{Temperature of the optics} \\
       \texttt{T\_contamination} & 1.000 & \parbox{10cm}{Effective throughput due to contamination} \\
       \hline
       \multicolumn{3}{c}{Coronagraph} \\
       \hline
       \\[-8pt]
       \texttt{Istar} & $3.076 \times 10^{-16}$ & \parbox{10cm}{Stellar intensity scalar ($I_{\mathrm{star}}$, on-axis PSF), product of the noise floor contrast of the coronagraph (uniform over dark hole) and the peak value of the off-axis PSF.} \\
       \texttt{noisefloor} & $1.025 \times 10^{-17}$ & \parbox{10cm}{Noise floor (NF) of the coronagraph} \\
       \texttt{skytrans} & 0.798 & \parbox{10cm}{Coronagraph's performance when observing an infinitely extended source} \\
       \texttt{coronagraph\_optical\_throughput} & 0.438 & \parbox{10cm}{Throughput of all coronagraphic optics} \\
       \hline
       \multicolumn{3}{c}{Detector} \\
       \hline
       \\[-8pt]
       \texttt{DC} [$e^{-}$ pixel$^{-1}$ s$^{-1}$] & $3.000 \times 10^{-5}$ & \parbox{10cm}{Dark current} \\
       \texttt{RN} [$e^{-}$ pixel$^{-1}$ read$^{-1}$] & 0 & \parbox{10cm}{Read Noise} \\
       \texttt{tread} [s] & 1000 & \parbox{10cm}{Time between reads} \\
       \texttt{CIC} [$e^{-}$ pixel$^{-1}$ ph$^{-1}$] & 0 & \parbox{10cm}{Clock-induced charge} \\
       \texttt{QE} & 0.750 & \parbox{10cm}{Quantum efficiency} \\
       \texttt{dQE} & 0.750 & \parbox{10cm}{Effective QE due to degradation} \\
       \hline
    \end{tabular}
    \caption{Parameters used in the \pyedith\ exposure time calculations for the EAC1 observatory preset. All noise instances are generated for a constant exposure time of $t_{\mathrm{exp}} = 10$ hours, and subsequently re-scaled as needed. Where units are not included, parameters are dimensionless.}
    \label{tab:pyedith}
\end{table*}

\subsection{Analytic Relation for Calculating Detectability as a Function of Resolution}

We derive an analytic relation for the minimum exposure time required to confidently detect a specific atmospheric gas in an exoplanet spectrum. {We note that detection is not the same as quantification, and that detectability thresholds do not translate to meaningful constraints on gas abundances.} Our approach uses the Bayesian Information Criterion (BIC) to compare two nested models: one including all gases (model 2) and one excluding the target gas (model 1).

Starting from the standard BIC formulation, where $k = 14$ is the number of model parameters for all cases and $N$ is the number of spectral data points, we require $\Delta$BIC$\geq10$ as our threshold for a gas detection, corresponding to strong Bayesian evidence favoring the complete model. $N$ changes depending on the resolution under investigation. The standard $\chi^2$ metric quantifies data-model goodness-of-fit given the uncertainty, $\sigma$. The key insight is how $\chi^2$ scales with observation quality and in turn exposure time. If we assume that HWO observations are in the photon-noise limit, spectral precision scales as $\sigma_1$ $\propto$ $1/\sqrt{t}$, so $\chi^2$ is directly proportional to exposure time $t$. That is, a model that omits a key atmospheric species that is present in the data (model 1) will see a linear increase in $\chi^2$ (a decrease in goodness-of-fit) as exposure time increases. 

For model 2 (all gases included), we assume perfect model-data agreement in the absence of random noise, giving $\chi^2_2 = 0$. This means the ``data'' is effectively the spectrum with all gases included. Model 1 (missing the target gas) will show residuals proportional to that gas's spectral signature. By expressing uncertainties at a new exposure time $t_2$ relative to a nominal exposure time $t_1$, we have $\sigma_2^2 = \sigma_1^2 \times (t_1/t_2)$. Substituting into $\Delta$BIC and solving for $t_2$, we obtain:

\begin{equation}
    t_2 = \frac{\Delta{\mathrm{BIC}}_0 - k\ln(N) + (k - 1)\ln(N)}{\left(\sum\limits_{i=1}^{N} \frac{(\mathrm{data_i} - \mathrm{model}_{1,i})^2}{\sigma_{1,i}^2 \times t_1} - \sum\limits_{i=1}^{N} \frac{(\mathrm{data}_i - \mathrm{model}_{2,i})^2}{\sigma_{1,i}^2 \times t_1}\right)}.
\end{equation}

Finally, because $\chi_2^2 = 0$ the second sum vanishes, simplifying to:

\begin{equation}
    t_2 = \frac{\Delta{\mathrm{BIC}}_0 - k\ln(N) + (k - 1)\ln(N)}{\left(\sum\limits_{i=1}^{N} \frac{(\mathrm{data} - \mathrm{model}_1)^2}{\sigma_{1,i}^2 \times t_1}\right)}.
\end{equation}

This expression directly relates the minimum required exposure time to the spectral signature strength of the target gas, providing a computationally efficient way to assess gas detectability in the visible and near-IR wavelength regimes. Accordingly, we generate \pyedith noise instances for each resolution case at a nominal exposure time $t_1 = 10$ hours, and use our analytical relationship to solve for the re-scaled exposure time $t_2$ that yields a $\Delta$BIC $\geq10$ strong detection of the missing gas. The resulting exposure time $t_2$ represents the minimum length of observation required to make a significant detection of a key gas in the visible or near-IR wavelength range for a given Earth through time case. We recalculate the minimum exposure time for each gas in each wavelength range to reflect the relative significance of spectral absorption features spanning both wavelength regimes. This is particularly applicable to \ch{H2O}, which has broad absorption features in both the visible and the near-IR and \ch{O2}, which has comparatively narrow absorption features at 0.76 \microns in the visible and at 1.27 \microns in the near-IR.


\subsection{\rfast Retrieval Model, Setup, \& Cases}

Retrievals enhance our interpretations of the analytical calculations by demonstrating how spectral resolution affects our ability to \textit{constrain} the abundances of key gases. This is particularly important for gases that we do not anticipate to be detectable, as the analytical detectability calculations inherently cannot show the nuances of potential changes to abundance upper limits. When it comes to ruling out the presence of antibiosignature gases, an upper limit can be significant to contextualize the detection of key gases by confirming a biogenic origin. This holds especially true for gases like CO where unrealistically long exposure times are required for detections. Retrievals are the appropriate tool of choice for tracking changes to upper limits on gases with changes to instrument design \citep{krissansen2025wavelength}.

We perform retrievals with \rfast, a 1-D atmospheric retrieval code from \cite{robinson2023exploring} that explores parameter space using the \emcee Markov Chain Monte Carlo (MCMC) sampler \citep{foreman2013emcee}. \rfast enables the retrieval of atmospheric, planetary, and orbital parameters via uniform or Gaussian priors in either log or linear space. In this study, all 14 parameters are retrieved in $\log10$ space assuming uniform priors. We retrieve seven unknown gas partial pressures, including \ch{N2}, \ch{O2}, \ch{H2O}, \ch{O3}, \ch{CO2}, \ch{CH4}, and CO. For \ch{N2} {and} \ch{H2O}, we impose a $\log10$ uniform prior ranging from $10^{-2}$--$10^7$ Pa. For the remaining gases, \ch{O2}, \ch{O3}, {\ch{CO2}, \ch{CH4},} and \ch{CO}, which may be trace species, we adopt a broader $\log10$ uniform prior ranging from $10^{-12}$--$10^7$ Pa. We also retrieve surface albedo (prior range 0.01--1.0), planet radius (prior range $10^{-0.5}$--$10^{0.5}$ $R_{\oplus}$), planet mass (prior range $0.1$--$10$ $M_{\oplus}$), cloud deck thickness (prior range $1$--$10^7$ Pa), cloud-top pressure (prior range $1$--$10^7$ Pa), cloud extinction optical thickness (prior range $10^{-3}$--$10^3$, and global cloud fraction (prior range $10^{-3}$--1). To ensure clouds form only above the surface, cloud top pressures smaller than the total atmospheric pressure are prohibited. For all cases a gray surface albedo is assumed. 

We set up our retrievals with 100 walkers within \emcee, where each walker takes $100{,}000$ steps with a burn-in of 10,000 and a thinning parameter of 100 to compress retrieval outputs. To accelerate MCMC convergence, we initialize all walkers in a Gaussian ``ball'' around the true values, after which all walkers are allowed to explore the entire prior range. We verify that \emcee has converged by confirming the chain lengths for all parameters exceed $50 \times \tau$, where $\tau$ is the autocorrelation time \citep{foreman2013emcee}.

As in previous studies \citep{hall2023constraining, krissansen2025wavelength}, we retrieve gas abundances as partial pressures to avoid assuming which gas is the bulk atmospheric constituent. This method is comparable to imposing center log ratio priors for atmospheric abundances \citep{benneke2012atmospheric, damiano2021reflected, damiano2022reflected}. Since the total atmospheric pressure is the sum of the retrieved gas partial pressures, surface pressure itself is not explicitly retrieved. 

\begin{figure*}[htbp]
    \centering
        \subfigure[]{\includegraphics[width=0.48\textwidth]{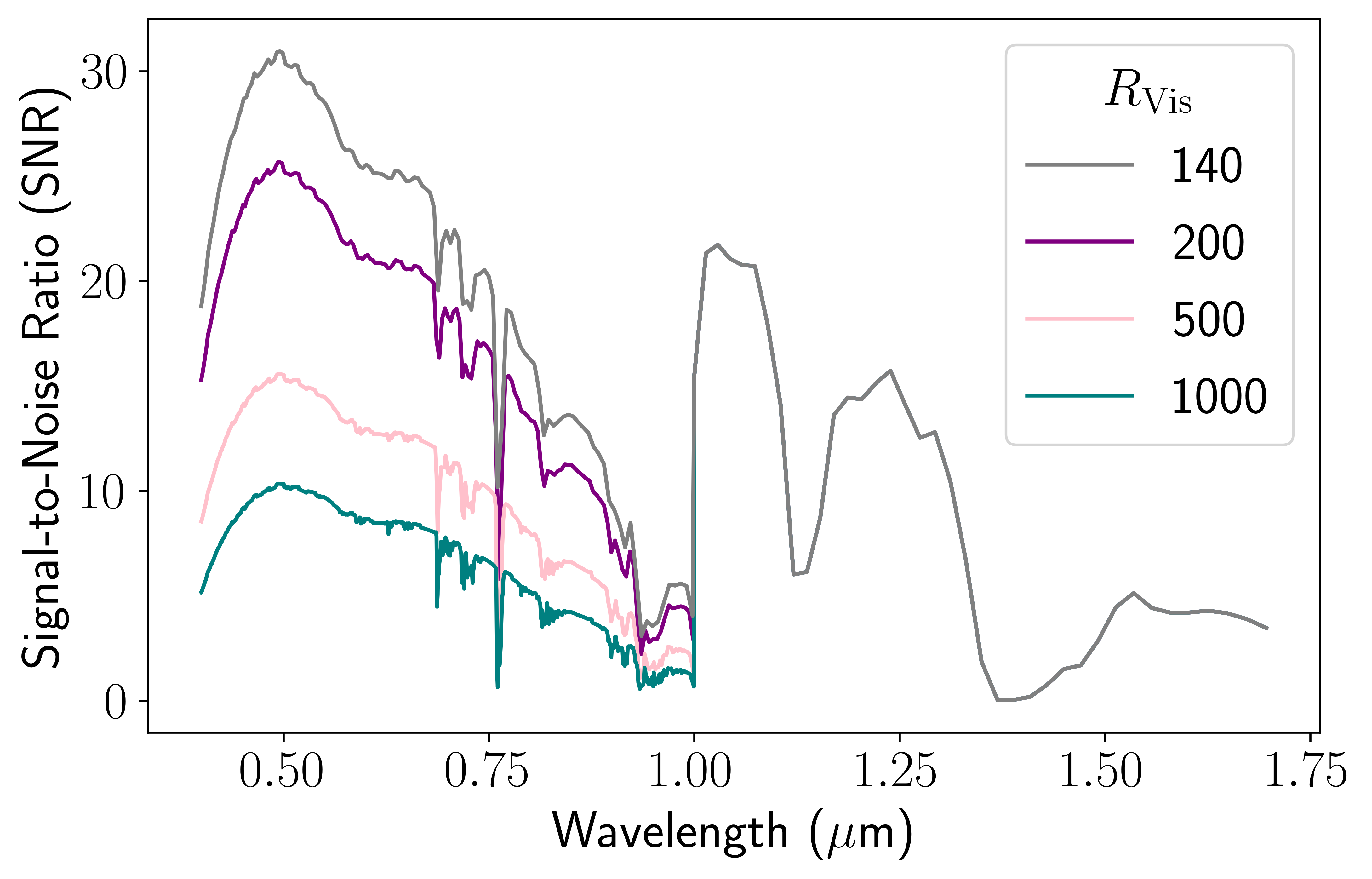}} 
        \subfigure[]{\includegraphics[width=0.48\textwidth]{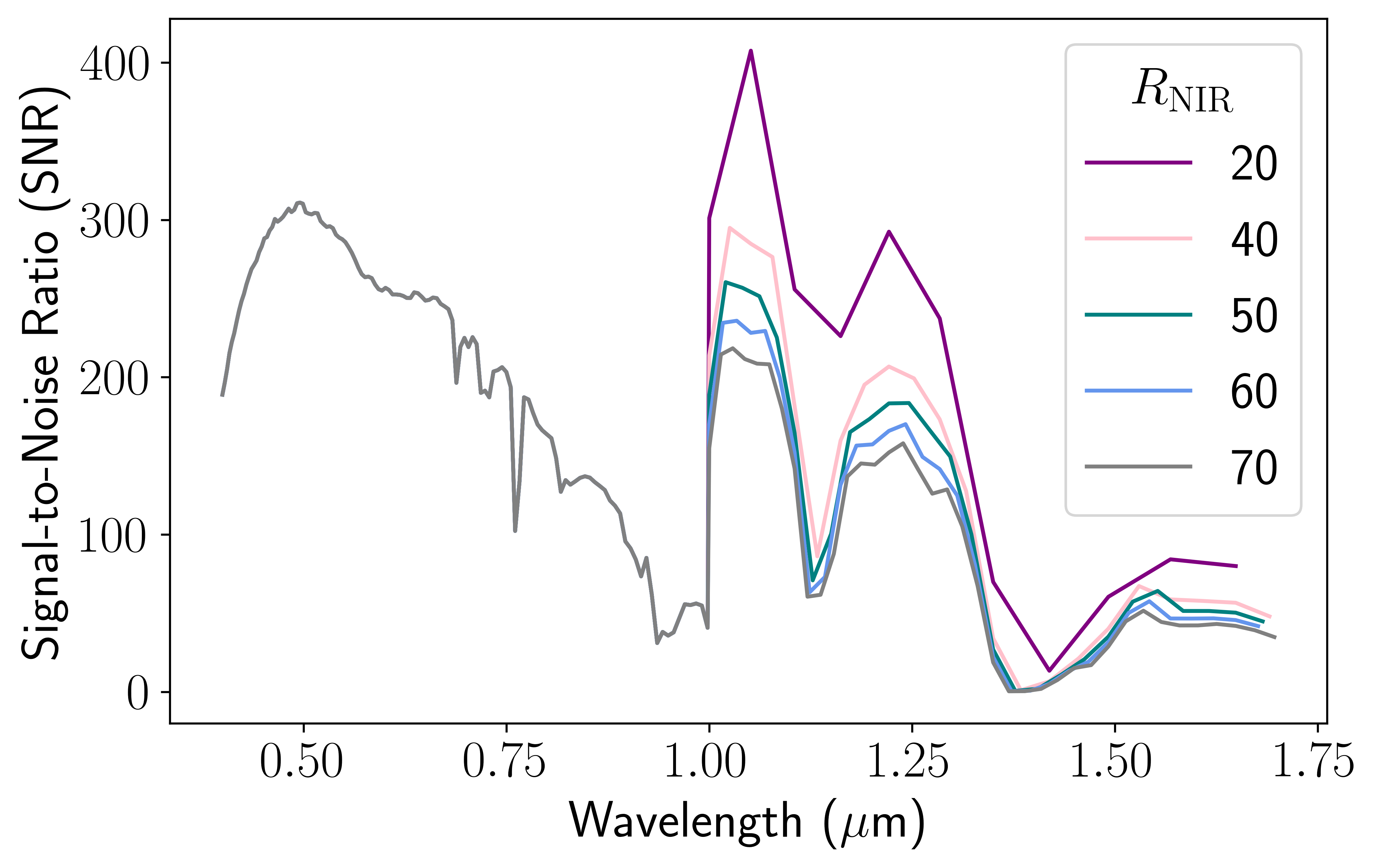}} 
    \caption{SNR as a function of wavelength for the visible and near-infrared resolution retrieval investigations using the exposure time calculator. When visible resolution is under investigation, the resolution in the near-infrared wavelengths is held at the nominal value (and vice versa when the near-infrared resolution is under investigation). Increased resolution leads to decreased SNR, while decreased resolution leads to increased SNR. Steep gradients in the signal-to-noise make it challenging to do an apples-to-apples comparison with prior works that use flat signal-to-noise. For example, for an SNR of 20 in the continuum near ${\sim}$1.1 \microns, the SNR at 1.5--1.75 \microns is $<$5.}
    \label{fig:SNR_lambda}
\end{figure*}

We summarize the retrieval cases included in this study in \autoref{tab:earth_cases}. We generated unique noise instances at every resolution case for a fixed exposure time of 10 hours. Then, we re-scaled the SNR of the entire spectrum by reducing the wavelength-dependent error bars by a constant factor. The constant factor was chosen to give near-band SNRs consistent with values in the literature. However, as shown in \autoref{fig:SNR_lambda}, the SNR varies significantly with wavelength due to the wavelength-dependent photon flux from the planet and instrument characteristics. This steep gradient in signal-to-noise makes direct comparisons with prior works that assume flat SNR challenging. For instance, an SNR of 20 in the continuum near ${\sim}$1.1 \microns corresponds to SNR ${\lesssim}$ 5 at 1.5--1.75 \microns (\autoref{fig:SNR_lambda}b). This wavelength dependence is crucial for understanding detectability across the full spectral range.


\begin{deluxetable*}{l@{\extracolsep{\fill}}cccc}
\setlength{\tabcolsep}{26pt}
\tablecaption{A summary of the retrieval cases explored in this study. For the visible-focused retrievals (0.4--1.0 \microns), we examine the Proterozoic and Phanerozoic cases with resolutions of 140, 200, 500, and 1000, while maintaining the near-IR resolution at the nominal value of 70 (in bold). For the near-IR-focused retrievals (1.0--1.7 \microns), we examine the late Archean (high \ch{CH4}) and Phanerozoic cases with resolutions of 20, 40, 50, 60, and 70, while maintaining the visible resolution at the nominal value of 140 (in bold). We include the SNR$_0$ at $\lambda_0$ used to re-scale the spectral uncertainties, and the corresponding constant exposure time.}
\tablehead{\colhead{Earth through} & \colhead{Resolutions} & \colhead{$\lambda_0$} & \colhead{SNR$_0$} & \colhead{Exposure} \\ [-0.3cm] \colhead{Time Case} & \colhead{($\lambda/\Delta\lambda$)} & \colhead{(\microns)} & \colhead{} & \colhead{Time (hours)}}
\startdata
\label{tab:earth_cases}
Archean & $R_{\mathrm{NIR}}$ = 20, 40, 50, 60, \textbf{70} & 1.6 & 40 & 8163 \\
\hline
Proterozoic & $R_{\mathrm{Vis}}$ = \textbf{140}, 200, 500, 1000 & 0.76 & 20 & 82 \\
\hline
\multirow{2}{*}{Phanerozoic} & $R_{\mathrm{Vis}}$ = \textbf{140}, 200, 500, 1000 & 0.76 & 20 & 82 \\
\cline{2-5}
& $R_{\mathrm{NIR}}$ = 20, 40, 50, 60, \textbf{70} & 1.6 & 40 & 8245 \\
\hline
\enddata
\end{deluxetable*}

\vspace{-24pt}

We perform retrievals on a set of spectra with differing resolving powers, but the same exposure time to isolate the information gain/loss strictly based on resolution changes. We use the nominal resolution case to determine exposure time scale factors that result in achieving some SNR$_0$ at $\lambda_0$ (see \autoref{tab:earth_cases}) based on past work targeting specific molecular bands that typically drive Earth-like characterization. For a given Earth-through-time case, once we determine the noise rescaling factor from the nominal resolution case ($R_{\mathrm{Vis}}=140$, $R_{\mathrm{NIR}}=70$), we apply this same scaling factor to all resolution cases. This approach isolates the physical effect of resolution changes on SNR (while maintaining identical exposure times between resolution cases): when resolution increases, photons are distributed across more spectral bins, reducing the SNR per bin; conversely, when resolution decreases, photons are concentrated into fewer bins, increasing the SNR per bin. 

For the cases where we investigate the effect of increasing visible resolution, we scale the SNR of the nominal resolution spectrum for each retrieval case so that there is an SNR $\geq$ 8.5 at the continuum of the \ch{O2} A-band feature at 0.76 \microns \citep{feng2018characterizing}. The re-scaled SNR$\sim$20 for the visible cases corresponds to a constant exposure time of $t_{\mathrm{exp}}$ ${\sim}$ 82 hours. For the cases where we investigate the effect of decreasing the resolution in the near-IR, we scale the SNR of the nominal resolution spectrum so that there is an SNR $\geq$ 20 in the continuum at the \ch{CO2} feature at ${\sim}$1.5\microns \citep{krissansen2025wavelength}. The re-scaled SNR$\sim$40 for the near-IR cases corresponds to a constant exposure time of $t_{\mathrm{exp}}$ ${\sim}$ 8200 hours. The substantially longer near-IR exposure times arise because the SNR per unit time is significantly lower in this wavelength range -- the planet reflects fewer near-IR photons, and detector thermal noise contributions are higher compared to visible wavelengths. 

For the visible wavelength investigations, we selected one Proterozoic case to determine whether increased spectral resolution reduces the exposure time required to detect \ch{O2} in low-\ch{O2} Earth-through-time atmospheres. We also examined the Phanerozoic case to establish a reference baseline, as modern Earth represents arguably the most compelling science case for HWO. For the near-IR investigations, we selected the high \ch{CH4} late Archean case, as the elevated abundances of \ch{CH4} and \ch{CO2} during this epoch are expected to produce stronger spectral features compared to the Proterozoic and Phanerozoic cases. We additionally conducted retrievals for the Phanerozoic case in the near-IR to provide a consistent baseline for comparison across wavelength regimes. The near-IR resolution investigation aims to assess whether reduced spectral resolution adversely impacts atmospheric characterization capabilities. Our analytic detectability calculations indicate that exposure time requirements for detecting key gases in the near-IR remain relatively insensitive to resolution degradation, suggesting potential engineering advantages for instrument development at lower near-IR resolving powers.

\begin{table}[h]
\setlength{\tabcolsep}{14pt}
\centering
\caption{Retrieval parameters and prior ranges. Pressure is not an independent free parameter as it is the sum of the constituent partial pressures (partial pressures are retrieved parameters). Parameters without explicit units reported are non-dimensional.}
\begin{tabular}{lc}
\toprule
Parameter & Prior range \\
\midrule
$p$\ch{N2}  & $[10^{-2}, 10^7]$ Pa \\
$p$\ch{O2} & $[10^{-12}, 10^7]$ Pa \\
$p$\ch{O3} & $[10^{-12}, 10^7]$ Pa \\
$p$\ch{CO2} & $[10^{{-12}}, 10^7]$ Pa \\
$p$\ch{CO} & $[10^{{-12}}, 10^7]$ Pa \\
$p$\ch{CH4} & $[10^{{-12}}, 10^7]$ Pa \\
$p$\ch{H2O} & $[10^{-2}, 10^7]$ Pa \\
Surface albedo ($A_s$) & $[0.01 - 1.0]$ \\
Radius ($R_p$) &  $[10^{-0.5} - 10^{0.5}]$ $R_{\oplus}$ \\
Mass ($M_p$) & $[10^{-1} - 10^{1}]$ $M_{\oplus}$\\
Cloud thickness ($\Delta p_c$) &  $[10^{0} - 10^{7}]$ Pa \\
Cloud top pressure ($p_t$) & $[10^{0} - 10^{7}]$ Pa \\
Cloud extinction ($\tau_c$) & $[10^{-3} - 10^{3}]$ \\
Cloud fraction ($f_c$) & $[10^{-3} - 10^{0}]$ \\
\bottomrule
\end{tabular}
\end{table}

\section{Results}
\label{sec:results}
We report the results of our investigation in the form of analytic detectability calculations and posterior distributions from atmospheric retrievals. For the former, we calculate the minimum exposure time required to make detections of key gases for our Earth through time cases as a function of resolution in the visible and near-IR wavelength regimes. We also assess how the magnitude of the dark current affects this minimum exposure time as a function of resolution. Finally, we report an acceptable range of instrument resolutions for detecting each gas in each Earth through time case using an exposure time threshold of 20\% from the minimum, which represent the size of the errors in our detectability calculations. For our retrieval cases investigating the effect of increased resolution in the visible wavelength regime, we analyze gas abundance posterior distributions retrieved for the high-\ch{CH4} Proterozoic case and the Phanerozoic Earth case. To assess the effect of decreased resolution in the near-IR, we show gas abundance posterior distributions retrieved for the high-\ch{CH4} Archean case and the Phanerozoic Earth case.

\subsection{Detectability as a Function of Resolution in the Visible and the Near-IR}

\begin{figure*}[htbp]
    \centering
    \includegraphics[width=\linewidth]{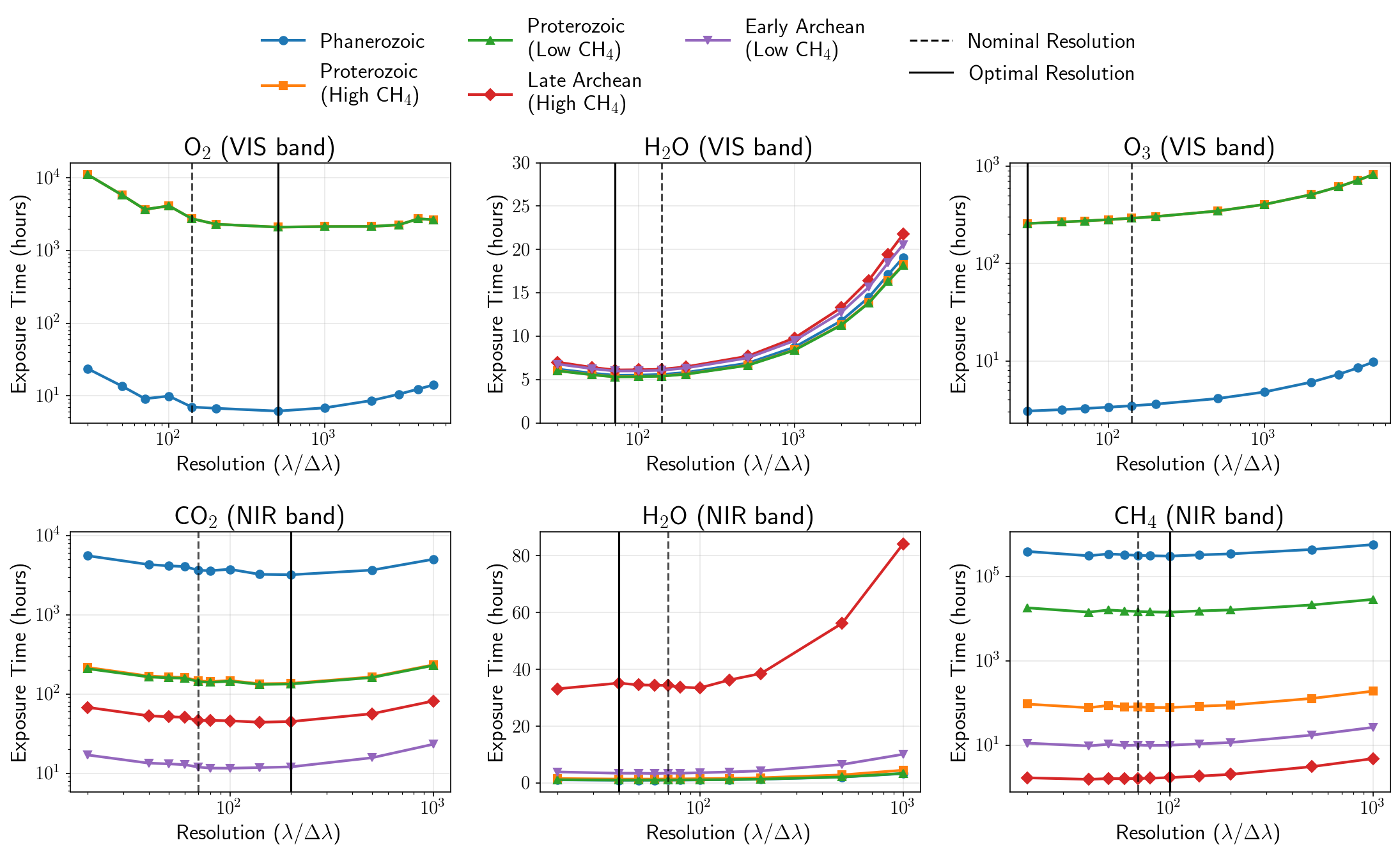}
    \caption{Exposure time required for gas detection as a function of resolution for key gases in corresponding band-passes for all Earth through time cases explored in this study, assuming the nominal dark current ($3 \times 10^{-5}$ e-/pixel/s) throughout. Here, ``VIS'' (top row) is shorthand for the visible band-pass, and ``NIR'' (bottom row) is shorthand for the near-infrared band-pass. The dashed vertical lines indicate the nominal HWO resolutions in the visible and the near-IR, while the solid vertical lines indicate the ``optimal'' resolution that gives the minimum exposure time according to our analytical calculations. Despite intrinsic band differences, our results show a generic trend of under-resolving bands where increased resolution decreases the exposure time, and over-resolving bands where increased resolution increases the exposure time by incurring excess pixel noise. In the visible, water, and ozone are broad and have a tendency to be over-resolved for most resolutions considered, whereas the more narrow-band \ch{O2} feature is more likely to be under-resolved.}
    \label{fig:exp_vs_res_sum}
\end{figure*}

\autoref{fig:exp_vs_res_sum} summarizes the relationship between exposure time and resolution for detecting key gases in relevant band-passes for all the Earth through time cases considered here. We indicate the nominal resolution with a vertical dotted line ($R_{\mathrm{Vis}}=140$ and $R_{\mathrm{NIR}}=70$ for the visible and near-IR, respectively), while we show our inferred ``optimal'' resolution, i.e. that which gives the minimal exposure time, with a solid vertical line. The top row shows the exposure time-resolution relationship for gases in the visible, including \ch{O2}, \ch{H2O}, and \ch{O3}. The bottom row shows the same relationship for gases in the near-infrared, including \ch{CO2}, \ch{H2O} and \ch{CH4}. The vertical offset between the different Earth through time cases occurs primarily due to differences in abundances. For example, detecting \ch{O2} and \ch{O3} in the visible for both Proterozoic atmospheres requires exposure times several orders of magnitude longer than for the Phanerozoic due to the extremely low oxygen abundances in these atmospheres. Similarly, the exposure time required to detect \ch{H2O} in the near-IR is an order of magnitude larger for the late Archean case, as the higher \ch{CH4} abundances lead to strong, overlapping absorption features with broad water vapor bands. 

Across all Earth through time cases, the gases in the visible and the near-IR demonstrate a similar pattern for exposure time as a function of spectral resolution. We find three resolution regimes that are associated with under-resolving molecular bands, optimally matching the resolution to the width of the bands, and over-resolving the bands. At insufficiently low resolutions, absorption bands are under-resolved and weakened by blending with the continuum, which requires additional exposure time to detect a given gas. At excessively high resolutions, absorption bands are over-resolved and many resolution elements (and pixels) span their widths, which also requires additional exposure time to build up the necessary signal to noise to detect a given gas due to per-pixel detector noise. We define the ``optimal resolution'' as that which balances these two effects and minimizes the exposure time. We note the exposure time-resolution relationship for gases in the near-IR is relatively flat in the vicinity of the nominal resolution, and exposure time changes relatively little when resolution is slightly degraded or enhanced. 

 We identify the optimal resolutions for each gas and band-pass via the solid vertical lines in \autoref{fig:exp_vs_res_sum}. Thus, our results suggest that shorter exposure times could be achieved for detecting \ch{O2} in all Earth through time cases by increasing resolution in the visible wavelengths to $R_{\mathrm{Vis}}=500$. By the same token, \ch{H2O} and \ch{O3} appear to be over-resolved in the visible bands, as the nominal resolution exceeds the optimal resolution. However, this is not surprising, given that the resolution in the visible is set not by the detection of these gases, but by the detection of the narrow-band \ch{O2} feature. In the near-IR, our results suggest that shorter exposure times could be achieved for detecting \ch{CO2} and \ch{CH4} by increasing the spectral resolution to $R_{\mathrm{NIR}}=200$ and $R_{\mathrm{NIR}}=100$, respectively. Similar to the visible wavelengths, \ch{H2O} remains slightly over-resolved. 

While the summary plots in \autoref{fig:exp_vs_res_sum} appear to show that the ``optimal'' resolution does not overlap with the current nominal instrument resolutions, point estimates of a single optimal resolution are inherently limited. Realistically, future missions must remain capable across a broad range of possible atmospheres, molecular species, and science cases that we are not fully considering here, but may need to be balanced during the design stage. Resolution ranges that deliver performance within a reasonable threshold of the minimum exposure time are therefore more meaningful and easier to incorporate in downstream design decisions than any single optimum value. Accounting also for the 20\% uncertainty in the signal-to-noise scaling calculation, we identify the practical range of resolutions required to detect key gases within a 20\% threshold of the minimum possible exposure time in hours. We show these calculated ranges for all Earth through time cases in \autoref{tab:exposure}, with key gases broken down by band-pass.  When we account for uncertainty in the table, nearly all acceptable resolution ranges encompass the nominal visible and near-IR resolutions. Our results suggest that only for detecting \ch{O2} in the Proterozoic would it be advantageous to go to slightly higher resolutions in the visible of 171, which is about 20\% higher than the nominal visible resolution of 140. However, we find that these gains are likely minimal since they still require very long exposure times of ${\sim}2000$ hours. In the near-IR, our results suggest that resolutions lower than nominal may also be sufficient for characterizing the Earth through time, particularly for Archean-like \ch{CH4} and \ch{CO2} abundances. Overall, our detectability calculation results suggest that the nominal resolutions in both the visible and the near-IR are likely suitable for detecting key gases. Moreover, our calculations suggest that these resolutions need not be strictly prescriptive -- there is some flexibility in the tradeoff between resolution and exposure time, and the nominal resolutions may be considered approximate targets within our provided ranges for instrument design considerations seeking to deliver the most scientifically capable telescope.

\begin{deluxetable*}{|c|c|c|c|c|c|}
\setlength{\tabcolsep}{14pt}
\tablecaption{For each Earth through time case, we show the permitted resolution ranges for detecting key gases within a 20\% threshold of the minimum possible exposure time in hours, assuming the nominal dark current ({3.0} $\times 10^{-5}$ e-/pixel/s) throughout. The nominal resolution is 140 in the visible (0.4-1.0 $\mu$m) and 70 in the near-IR (1.0-1.7 $\mu$m). In the ``interpretation'' column, we indicate for each gas whether the nominal resolution falls in the acceptable range, and whether the corresponding minimum exposure time is reasonable ($t < 1000$ hours). ``SR'' means that the nominal resolution is sufficient and yields a reasonable exposure time, ``SU'' means that the nominal resolution is sufficient but yields an unrealistically long exposure time, and ``ISU'' means that the nominal resolution is insufficient and the required resolution range yields an unrealistically long exposure time. {Note that we do not find that altering the nominal resolution in the visible or near-IR wavelength ranges would significantly shorten the exposure times required to detect any of the gases across the Earth through time cases studied here.} \label{tab:exposure}}
\tablewidth{0pt}
\tablehead{
\colhead{Earth through} & \colhead{} & \colhead{} & 
\colhead{Min. Exposure} & \colhead{Resolution} & \colhead{} \\ [-0.3cm]
\colhead{Time Case} & \colhead{Gas} & \colhead{Bandpass} & 
\colhead{Time (hours)} & \colhead{Range ($\lambda$/$\Delta\lambda$)} & \colhead{Interpretation}} 
\startdata
\multirow{5}{*}{Early Archean (Low \ch{CH4})} & \multirow{2}{*}{\ch{H2O}} & Visible & 5.98 & 30-420 & SR \\
\cline{3-6}
 & & Near-IR & 3.38 & 20-164 & SR \\
\cline{2-6}
 & \ch{CO2} & Near-IR & 11.56 & 38-352 & SR \\
\cline{2-6}
 & CO & Near-IR & $5.5\times10^8$ & 197-782 & SU \\
\cline{2-6}
 & \ch{CH4} & Near-IR & 9.61 & 20-195 & SR \\
\hline
\multirow{5}{*}{Late Archean (High \ch{CH4})} & \multirow{2}{*}{\ch{H2O}} & Visible & 6.10 & 30-410 & SR \\
\cline{3-6}
 & & Near-IR & 33.16 & 20-221 & SR \\
\cline{2-6}
 & \ch{CO2} & Near-IR & 43.99 & 42-406 & SR \\
\cline{2-6}
 & CO & Near-IR & $1.8\times10^{6}$ & 57-706 & SU \\
\cline{2-6}
 & \ch{CH4} & Near-IR & 1.55 & 20-142 & SR \\
\hline
\multirow{9}{*}{Proterozoic (Low \ch{CH4})} & \multirow{2}{*}{\ch{H2O}} & Visible & 5.30 & 30-422 & SR \\
\cline{3-6}
 & & Near-IR & 1.04 & 20-144 & SR \\
\cline{2-6}
 & \multirow{2}{*}{\ch{O2}} & Visible & 2099.93 &  171-3548 & ISU \\
\cline{3-6}
 & & Near-IR & $1.3\times10^{6}$ & 60-339 & SU \\
\cline{2-6}
 & \ch{O3} & Visible & 257.27 & 30-242 & SR \\
\cline{2-6}
 & \ch{CO2} & Near-IR & 132.13 & 60-466 & SR \\
\cline{2-6}
 & CO & Near-IR & 1.1e+13 & 58-407 & SU \\
\cline{2-6}
 & \ch{CH4} & Near-IR & $1.4\times10^4$ & 26-256 & SU \\
\hline
\multirow{9}{*}{Proterozoic (High \ch{CH4})} & \multirow{2}{*}{\ch{H2O}} & Visible & 5.32 & 30-421 & SR \\
\cline{3-6}
 & & Near-IR & 1.41 & 20-147 & SR \\
\cline{2-6}
 & \multirow{2}{*}{\ch{O2}} & Visible & 2100.57 &  171-3549 & ISU \\
\cline{3-6}
 & & Near-IR & $1.3\times10^{6}$ & 60-339 & SU \\
\cline{2-6}
 & \ch{O3} & Visible & 257.30 & 30-242 & SR \\
\cline{2-6}
 & \ch{CO2} & Near-IR & 134.60 & 61-468 & SR \\
\cline{2-6}
 & CO & Near-IR & $1.1\times10^{13}$ & 58-407 & SU \\
\cline{2-6}
 & \ch{CH4} & Near-IR & 77.34 & 22-225 & SR \\
\hline
\multirow{9}{*}{Phanerozoic} & \multirow{2}{*}{\ch{H2O}} & Visible & 5.49 & 30-416 & SR \\
\cline{3-6}
 & & Near-IR & 1.00 & 20-140 & SR \\
\cline{2-6}
 & \multirow{2}{*}{\ch{O2}} & Visible & 6.17 & 135-1322 & SR \\
\cline{3-6}
 & & Near-IR & 262.85 & 50-299 & SR \\
\cline{2-6}
 & \ch{O3} & Visible & 3.06 & 30-241 & SR \\
\cline{2-6}
 & \ch{CO2} & Near-IR & 3248.13 & 66-564 & SU \\
\cline{2-6}
 & CO & Near-IR & $5.9\times10^{12}$ & 60-366 & SU \\
\cline{2-6}
 & \ch{CH4} & Near-IR & $3.1\times10^{5}$ & 27-269 & SU \\
\hline
\enddata
\end{deluxetable*}

\vspace{-24pt}

\begin{figure*}
    \centering

    \textbf{(a)}\\[0.3em]
    \begin{minipage}{0.32\textwidth}\centering
        \small(i)\\
        \includegraphics[width=\linewidth]{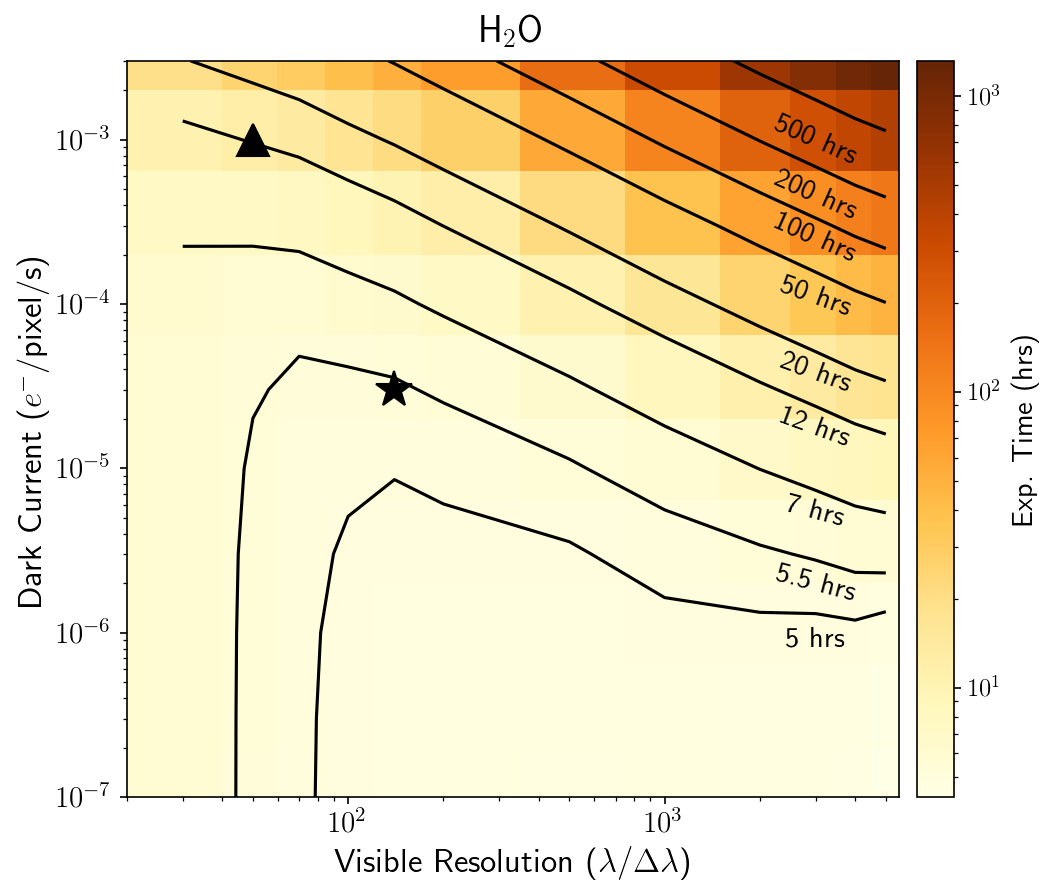}
    \end{minipage}\hfill
    \begin{minipage}{0.32\textwidth}\centering
        \small(ii)\\
        \includegraphics[width=\linewidth]{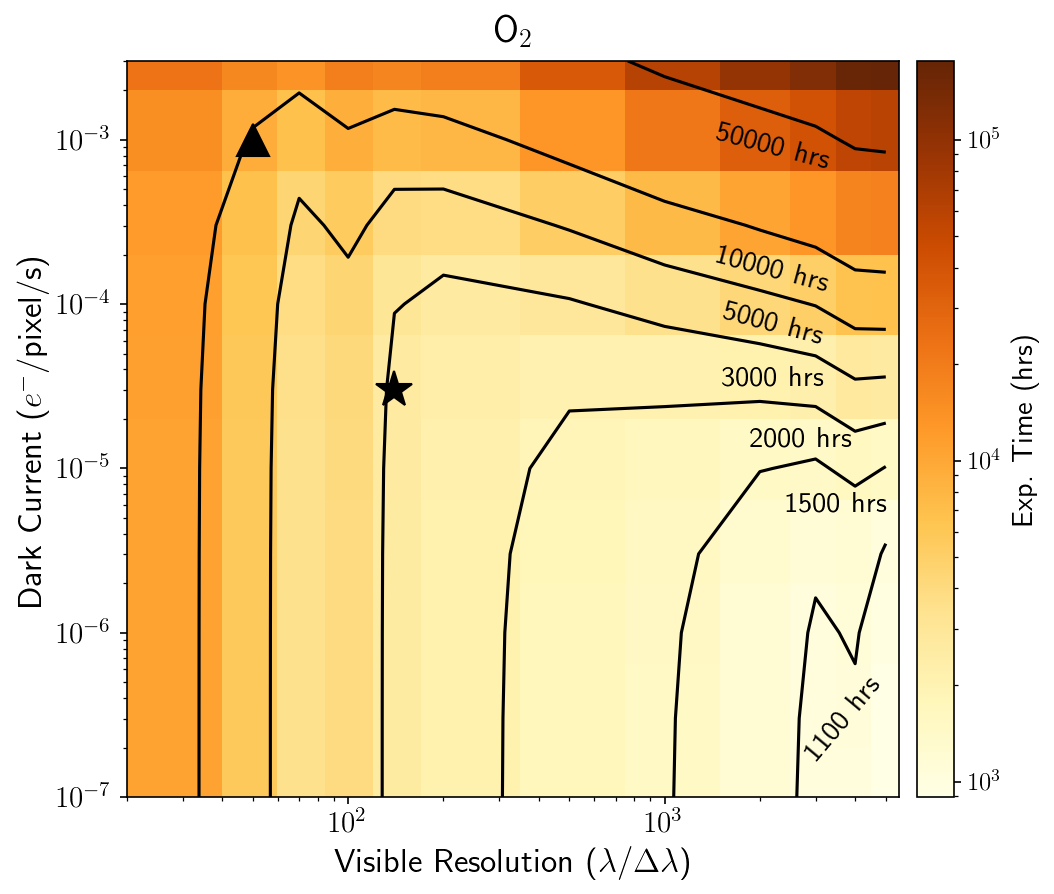}
    \end{minipage}\hfill
    \begin{minipage}{0.32\textwidth}\centering
        \small(iii)\\
        \includegraphics[width=\linewidth]{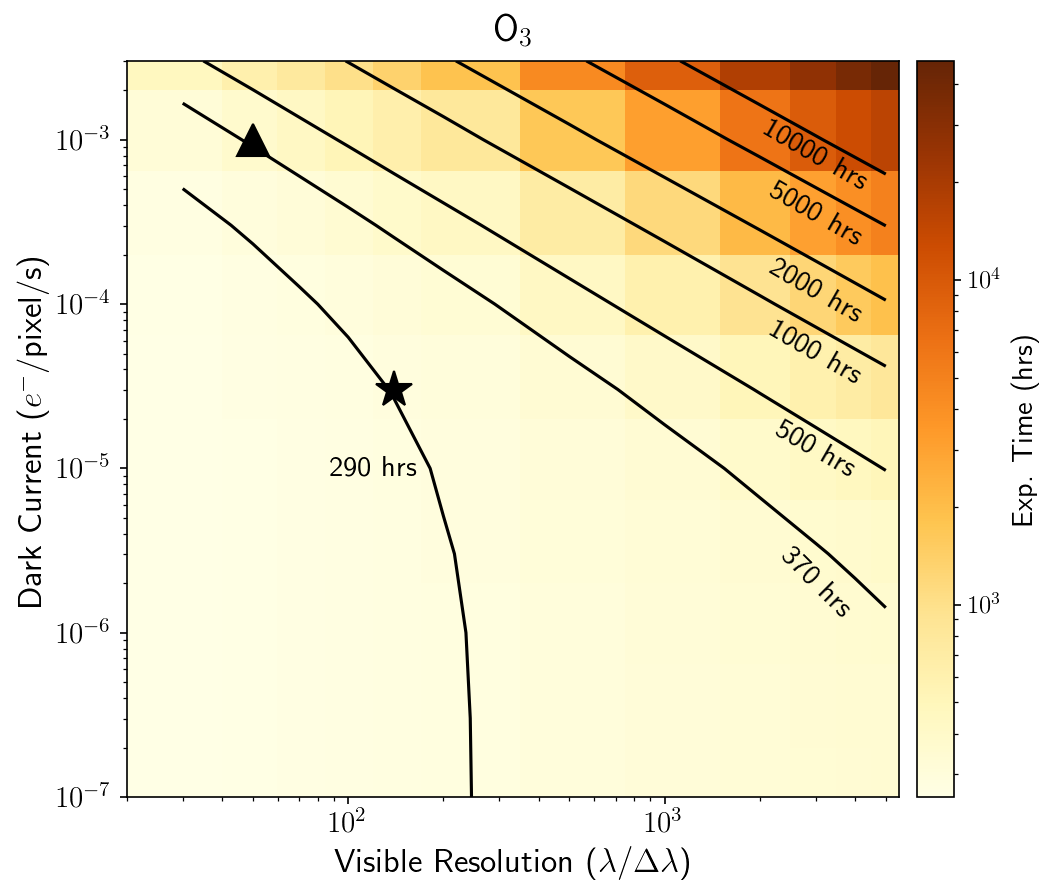}
    \end{minipage}

    \vspace{1em}

    \textbf{(b)}\\[0.3em]
    \begin{minipage}{0.32\textwidth}\centering
        \small(i)\\
        \includegraphics[width=\linewidth]{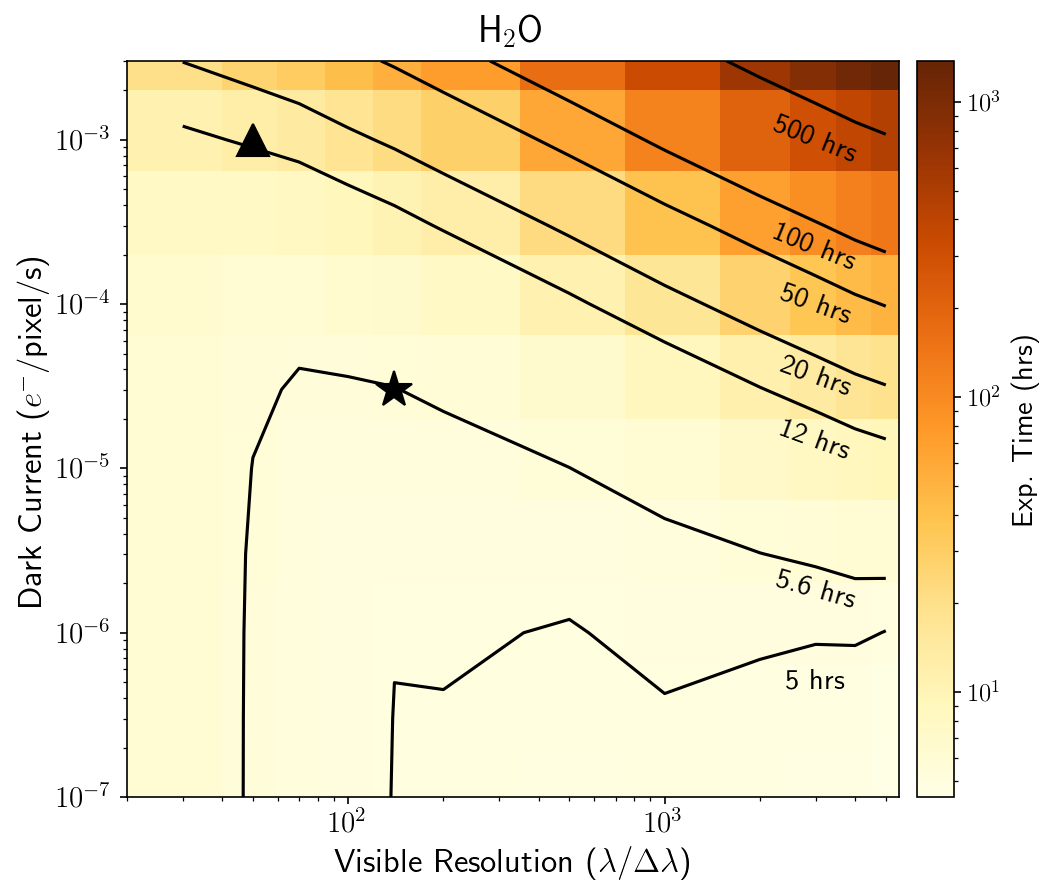}
    \end{minipage}\hfill
    \begin{minipage}{0.32\textwidth}\centering
        \small(ii)\\
        \includegraphics[width=\linewidth]{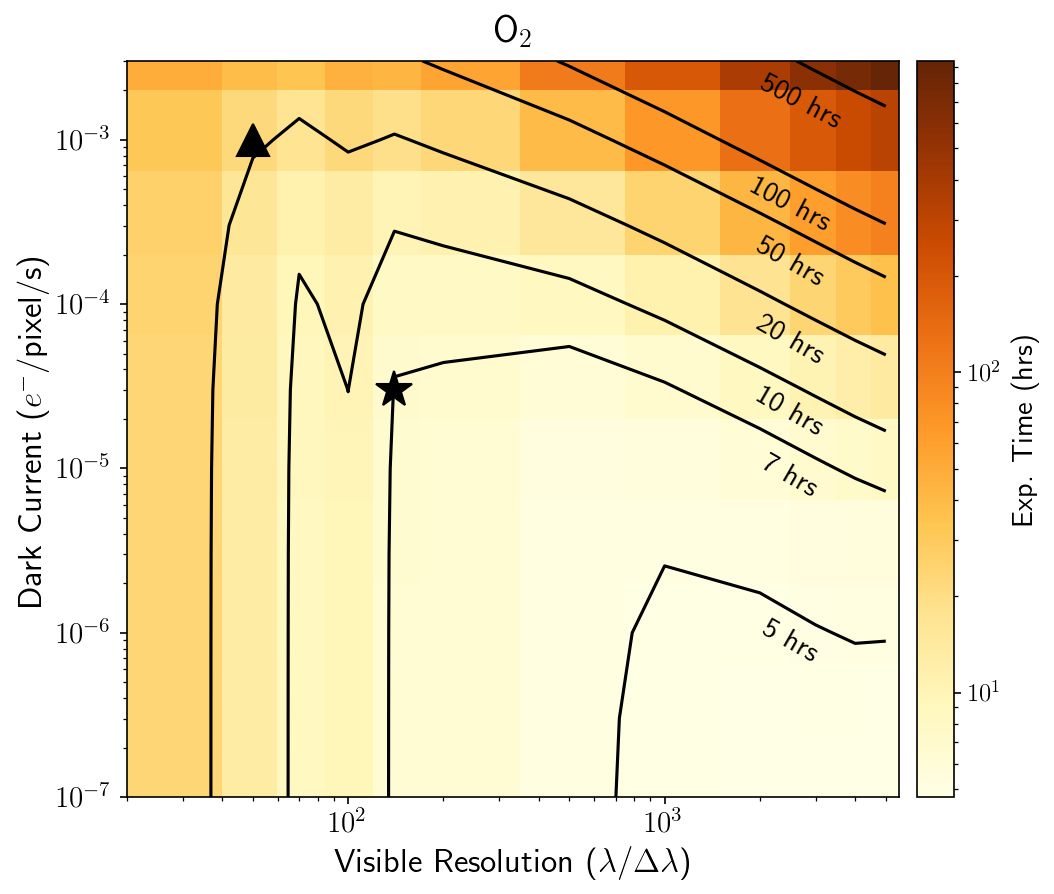}
    \end{minipage}\hfill
    \begin{minipage}{0.32\textwidth}\centering
        \small(iii)\\
        \includegraphics[width=\linewidth]{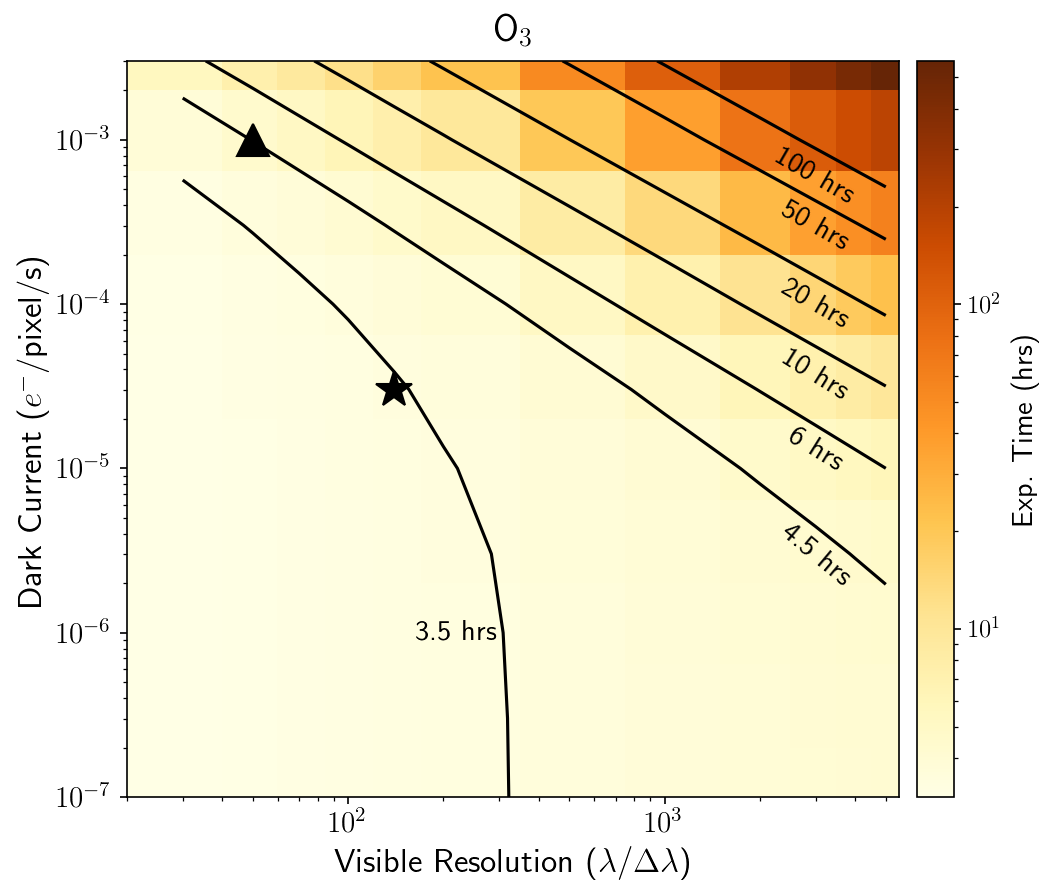}
    \end{minipage}

    \caption{Contour plots showing the relationship between dark current, visible band-pass resolution, and exposure time required for molecular detection. \textbf{(a)} High-\ch{CH4} Proterozoic Earth: (i) \ch{H2O}, (ii) \ch{O2}, (iii) \ch{O3}. While higher visible resolution correlates with lower relative exposure times to detect \ch{O2} at low Proterozoic abundances, detection remains infeasible in a reasonable time for all cases. \textbf{(b)} Phanerozoic Earth: (i) \ch{H2O}, (ii) \ch{O2}, (iii) \ch{O3}. Increasing visible resolution could lower \ch{O2} detection exposure times, but only if detector noise is also significantly reduced from the nominal value. In both panels, the star marks nominal HWO EAC1 values in \pyedith, and the triangle marks the nominal resolution ($R{\sim}50$) and dark current ($1\times10^{-3}$ e$^-$/pixel/s) of the Nancy Grace Roman Telescope Coronagraph Instrument Spectrograph.}
    \label{fig:contour-combined}
\end{figure*}

Our detectability calculation results shown in \autoref{fig:exp_vs_res_sum} and \autoref{tab:exposure} assume a nominal dark current value of $3 \times 10^{-5}$ $e^-$/pixel/s. \autoref{fig:contour-combined} demonstrates how varying dark current affects the relationship between resolution and exposure time in the visible band-pass for the high-\ch{CH4} Proterozoic case (a) and the Phanerozoic case (b), respectively. For each case, we show the effect on the detectability of \ch{H2O} (i), \ch{O2} (ii), and \ch{O3} (iii). On each subplot, the star denotes the case corresponding to the nominal dark current and nominal visible resolution ($R_{\mathrm{Vis}} = 140$), and the contours show combinations of dark current and resolution that give equivalent exposure times. We also show the noise ($1 \times 10^{-3}$ $e^-$/pixel/s) and spectral resolution ($R=50$) of the Nancy Grace Roman Telescope's Coronagraph Instrument (CGI), demarcated as a triangle on the plots. CGI will also provide spectroscopic observations in the visible wavelengths, providing a useful baseline for comparison of current and future coronagraph technologies. 

The non-smooth features visible in the contour plots (e.g., at $R_{\mathrm{Vis}} \approx 100$ and $R_{\mathrm{Vis}} \approx 3000$--4000) are numerical artifacts arising from pixel-phase aliasing on the discrete ETC wavelength grid, wherein narrow absorption band cores are comparable in width to individual pixels and $\Delta\chi^2$ varies non-monotonically depending on whether pixel centers happen to align with the deepest part of the absorption features as resolution changes. We confirm this interpretation by rigidly shifting the ETC wavelength grid by sub-pixel increments and recomputing $\Delta\chi^2$, finding that the resulting variability in detection metrics is largest where contour irregularities appear and smallest where contours are smooth, consistent with pixel-phase aliasing rather than any physical discontinuity in detectability. More broadly, this sensitivity suggests that for high-priority narrow-band features such as \ch{O2} that are at risk of being under-resolved, instrument design teams should exercise caution to ensure that resolution elements are centered on the band core rather than straddling it, as misalignment between pixel boundaries and narrow absorption features could meaningfully degrade detection significance.

\autoref{fig:contour-combined} demonstrates that HWO's projected dark current represents over an order of magnitude improvement over CGI. For extremely low-noise detectors with dark current below $1 \times 10^{-6}$ $e^-$/pixel/s, higher resolutions grant gas detections in shorter exposure times. More generally, there is a resolution threshold above which dark current must be suppressed to avoid accruing increased exposure time. While higher resolution in the visible reduces the required exposure time to detect \ch{O2} in low-\ch{O2} Proterozoic atmospheres, the minimum exposure time is still on the order of thousands of hours. Thus, any gains from increased spectral resolution in the optical are likely irrelevant based on this case alone. For modern-Earth \ch{O2} in the Phanerozoic, decreasing the visible resolution from the nominal would rapidly increase exposure times regardless of dark current, while increasing resolution would yield comparable exposure times up to $R_{\mathrm{Vis}} \approx 1000$, after which lower dark current detectors would be required to prevent exposure time increases. However, because \ch{H2O} is already over-resolved at the nominal resolution, increasing the resolution up to $R_{\mathrm{Vis}} \approx 1000$ would increase exposure times to detect \ch{H2O} by ${\sim}50\%$ unless detector noise is decreased by nearly an order of magnitude. Thus, an optimal instrument design for detecting one molecule may be suboptimal for another molecule.

\subsection{Retrieval Cases}

We followed up on the results of our analytical exploration of resolution and exposure time by performing retrievals for the most interesting and potentially requirement-driving cases.  The implication that higher resolution in the visible may reduce the exposure time required to detect \ch{O2} is particularly compelling in the case of the low-\ch{O2} Proterozoic Earth. Conversely, we further assess the implication that going to lower resolution in the near-infrared would not significantly increase the required exposure time in the case of the Archean Earth. For both the visible and the near-infrared, we also performed retrievals for the Phanerozoic Earth, as this is an important nominal case that will guide HWO instrument design. {Full corner plots containing the marginal distributions and covariances of all retrieved parameters are shown in the Appendix, along with a detailed discussion of trends in retrieved cloud parameters.} 

\begin{figure*}[htbp]
    \centering

   \noindent\textbf{(a)}\\
    \includegraphics[width=\linewidth]{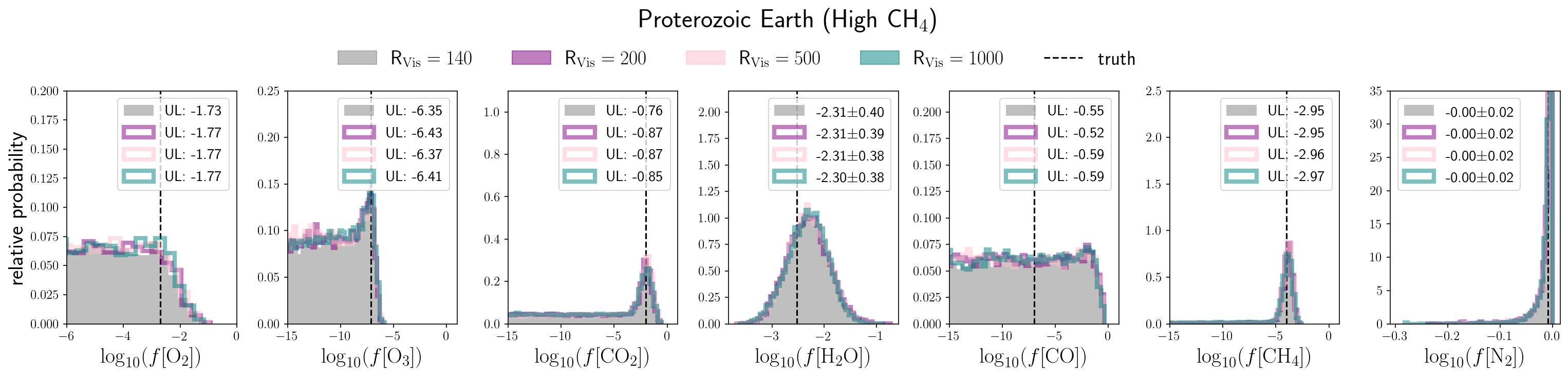}\\[0.3em]
       \noindent\textbf{(b)}\\
    \includegraphics[width=\linewidth]{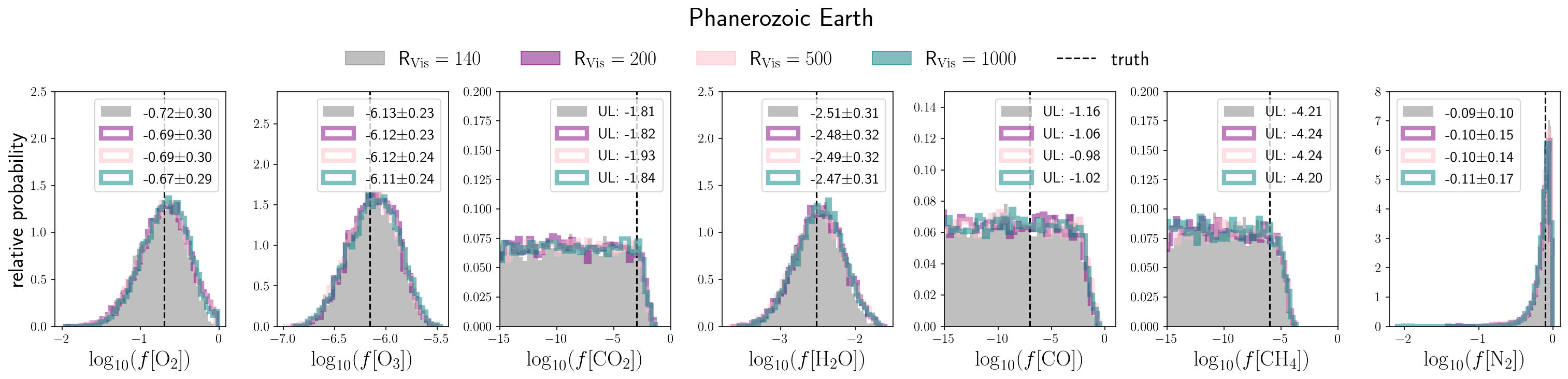}\\[0.3em]
       \noindent\textbf{(c)}\\
    \includegraphics[width=\linewidth]{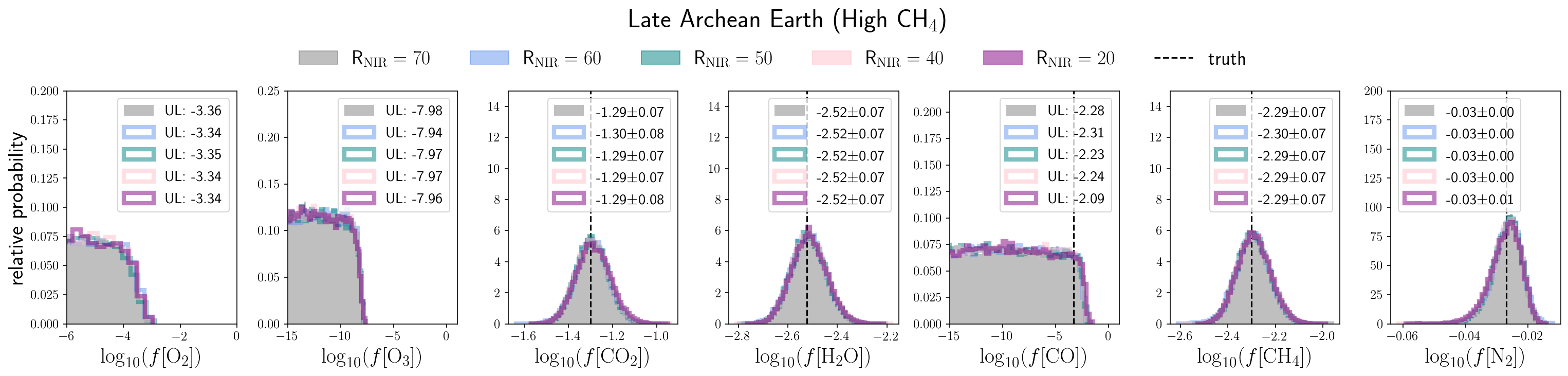}\\[0.3em]
       \noindent\textbf{(d)}\\
    \includegraphics[width=\linewidth]{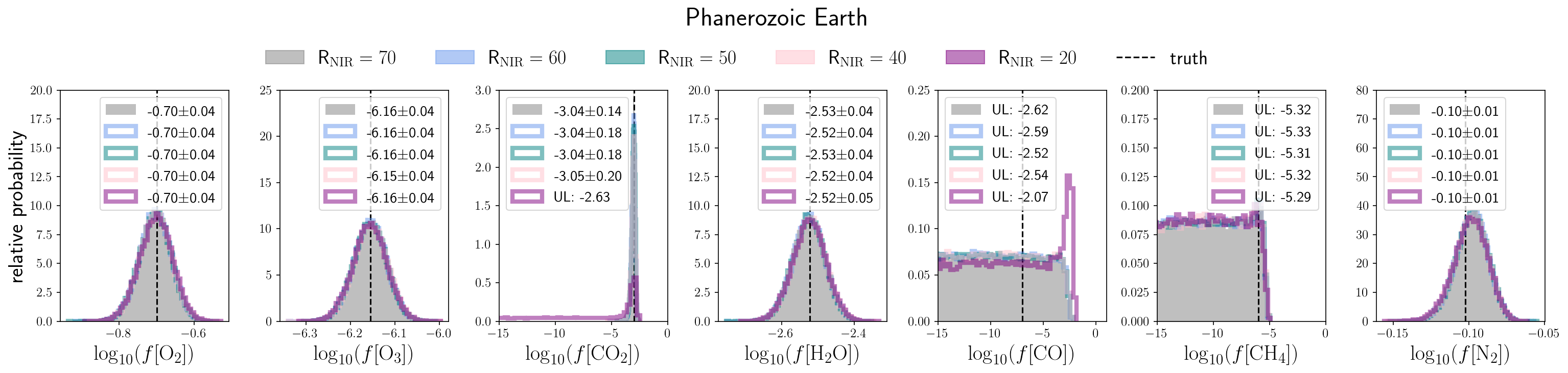}

    \caption{Retrieval results with varying spectral resolution across Earth analog scenarios. \textbf{(a)} Proterozoic Earth, visible band-pass (NIR held at $R=70$, $t = 81.63$ hrs). \textbf{(b)} Phanerozoic Earth, visible band-pass (NIR held at $R=70$, $t = 81.63$ hrs). In both (a) and (b), no advantage is conferred by higher visible resolution for \ch{O2}, \ch{O3}, and \ch{H2O} features. \textbf{(c)} Archean Earth, near-IR (VIS held at $R=140$, $t = 8163.27$ hrs); no significant advantage from higher NIR resolution for \ch{H2O}, \ch{CO2}, \ch{CH4}, and \ch{CO} features. \textbf{(d)} Phanerozoic Earth, near-IR (VIS held at $R=140$, $t = 8245.51$ hrs); degrading to $R_{\mathrm{NIR}}=20$ causes loss of \ch{CO2} detection and increased \ch{CO} upper limits due to blending of neighboring features.}
    \label{fig:resolution_retrievals}
\end{figure*}

\begin{figure}[htbp]
\centering
\includegraphics[width=\linewidth]{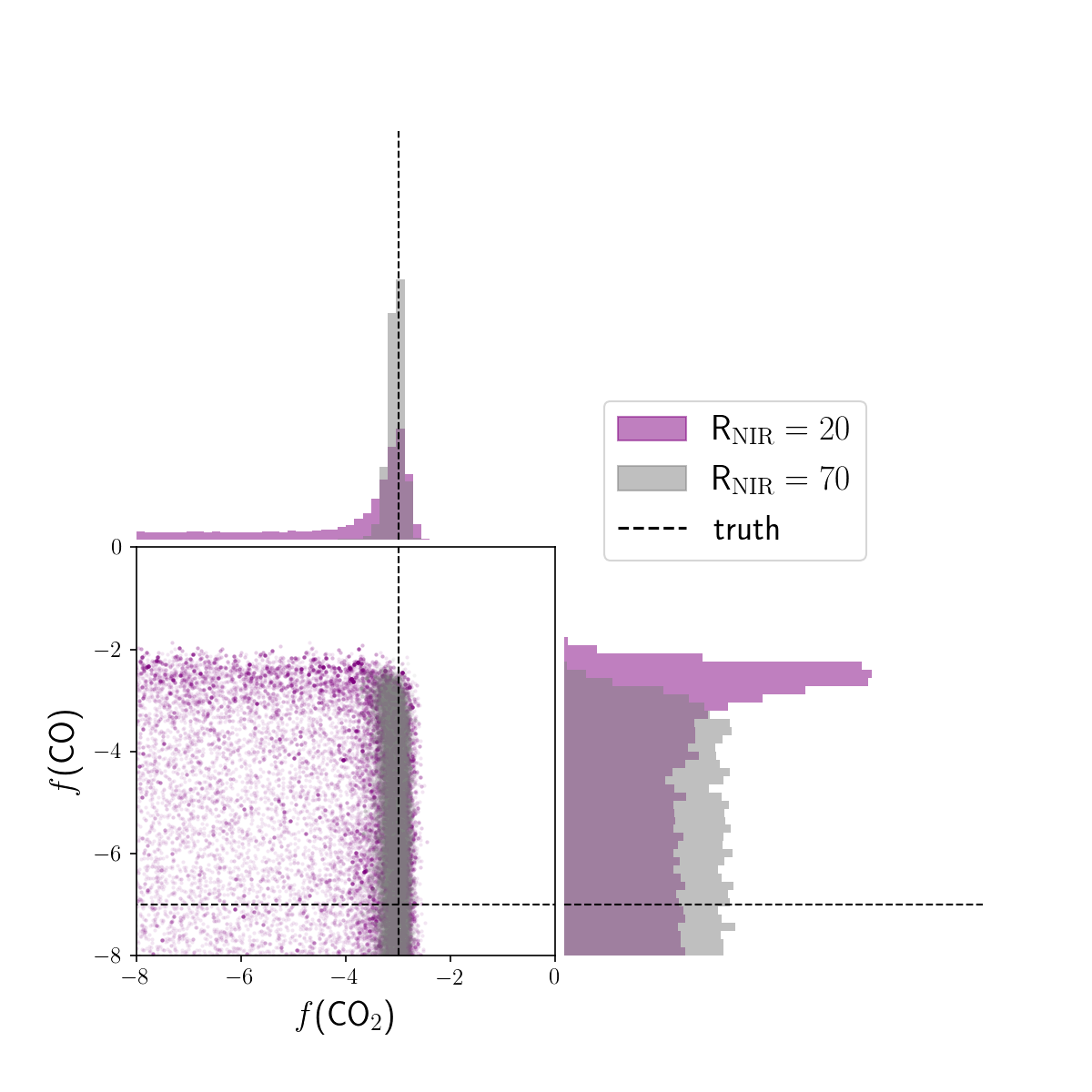}
\caption{A comparison of the marginal and covariance plots of the retrieved \ch{CO2} and \ch{CO} abundances for the Phanerozoic Earth, with $R_{\mathrm{NIR}} = 20$ and the nominal $R_{\mathrm{NIR}} = 70$. When the spectral resolution in the near-IR is lowered to 20, the retrieval results suggest that the data favors an inaccurately high abundance of CO combined with lower \ch{CO2} abundances.}
\label{fig:covar}
\end{figure}

We show retrieved posterior distributions for the cases of interest in \autoref{fig:resolution_retrievals}. All retrievals for a given case have a fixed exposure time to isolate how atmospheric constraints vary with spectral resolution in the wavelength region of interest. In each plot, we show the posterior distributions for all atmospheric constituents in terms of $\log10$ abundances, with different spectral resolutions shown as different colors. The retrieval corresponding to the nominal resolution is shown filled-in gray, while the remaining cases are shown as step-style histograms. In each subplot, a dashed vertical line represents the ``true'' input abundance of a given gas. Finally, a legend in each subplot indicates whether we classify the posteriors (by inspection) as (a) a non-detection upper limit (``UL'') with the 3-$\sigma$ limit given in log-scale, or (b) a detection constraint, with the median retrieved abundance provided with the 1-$\sigma$ standard deviation as an uncertainty, both in log-scale. 

For our investigation of spectral resolution in the visible wavelengths, our retrieval results for the Proterozoic (\autoref{fig:resolution_retrievals}a) and Phanerozoic Earth (\autoref{fig:resolution_retrievals}b) show slight variations in upper limits that are all within 0.18 dex. For \ch{O2} in the Proterozoic case, variations in upper limits are within $\pm0.04$ dex. For gas detections in both cases, slight variations in the median retrieved abundances are well within the $1\sigma$ uncertainties, with no significant differences in posterior widths. Ozone is not detected in the Proterozoic case due the exclusion of the UV wavelengths. For the Phanerozoic atmosphere, we obtain only upper limits on \ch{CO2}, \ch{CH4}, and \ch{CO}. This is because our constant exposure time of $t_{\mathrm{exp}}{\sim}80$ hours did not yield sufficiently high SNR in the near-IR wavelengths where these gases have significant absorption features.

Similar to the Proterozoic and Phanerozoic cases at visible wavelengths, our retrieval results in the near-infrared for the Archean (\autoref{fig:resolution_retrievals}c) show a relative insensitivity to lower spectral resolutions. Slight variations in upper limits are all within 0.09 dex, and median retrieved abundances overlap within $1\sigma$ uncertainties. However, our results for the Phanerozoic Earth (\autoref{fig:resolution_retrievals}d) show an interesting difference. For the Phanerozoic case, though near-infrared resolutions between 40 and 70 follow a similar pattern, the lowest resolution $R_{\mathrm{NIR}}=20$ shows a clear difference in the retrieved \ch{CO2} and \ch{CO} posterior distributions. 

At the lowest resolution that we explored in the near-infrared ($R_{\mathrm{NIR}}=20$), the retrieval results show that we lose confidence in the unambiguous detection of \ch{CO2} (as evidenced by the extended tail to low \ch{CO2} abundances). At the same time, the 3$\sigma$ upper limit on \ch{CO} increases by 0.55 dex relative to the nominal resolution case. Both posteriors imply a worsening ambiguity in the abundances of \ch{CO} and \ch{CO2}, where the data slightly favors an unrealistically high \ch{CO} abundance with correspondingly low \ch{CO2} abundances. This ambiguity arises because the extremely coarse resolution effectively blends the weak \ch{CO2} and \ch{CO} near-infrared features, resulting in a covariance favoring low \ch{CO2} abundances with high \ch{CO} abundances. To show the underlying degeneracy between \ch{CO} and \ch{CO2}, in \autoref{fig:covar} we compare the 1D and 2D marginalized posteriors of \ch{CO2} and \ch{CO} for the Phanerozoic case for both $R_{\mathrm{NIR}} = 20$ and the nominal $R_{\mathrm{NIR}} = 70$. The \ch{CO}-\ch{CO2} covariance shows that the long tail extending to lower \ch{CO2} abundances results in a corresponding probability build-up at higher \ch{CO} abundances, creating a visible peak in the \ch{CO} marginal distribution that could mislead interpretations.

While decreased resolution appears to have little to no effect on the retrieved posteriors for the Archean case, the Phanerozoic case demonstrates a clear disadvantage of excessively degrading the resolution in the near-infrared. This is likely because the Phanerozoic has lower abundances of \ch{CO2}, and thus weaker absorption features. At higher, Archean-like \ch{CO2} abundances, additional absorption features that do not overlap with \ch{CO} help break this degeneracy, allowing for a confident detection of \ch{CO2}. {Finally, cloud parameter retrievals are largely robust to changes in spectral resolution, though the lowest resolution case ($R_{\mathrm{NIR}}=20$) introduces both modest precision losses and a small but nonzero systematic bias in the cloud opacity (see Appendix).} 

\section{Discussion}
\label{sec:discussion}

Our study produces results in two forms: analytical detectability calculations, and retrieved posterior distributions. Together, these methods are complementary. The detectability calculations provide the minimum spectral resolution, dark current, and exposure time required to confidently detect gases via isolated absorption features, whereas retrievals demonstrate how resolution trade space affects our ability to accurately and meaningfully characterize the planetary atmosphere. Given that the detectability calculations are less nuanced, instrument parameters inferred from this analysis represent general guidance specific to the science cases investigated here. Conversely, retrievals provide the most rigorous assessment of how spectral resolution may impact the search for life on Earth-like exoplanets.  

In the visible wavelengths, we are primarily concerned with the spectral resolution required to detect and constrain \ch{O2}, \ch{O3}, and \ch{H2O}. While our detectability calculation results (e.g., \autoref{tab:exposure}) show that \ch{O3} and \ch{H2O} may be detected at lower resolving powers ($R_{\mathrm{Vis}}\ge30$), the spectral resolution in the visible should be set by the narrow \ch{O2}-A band, which requires higher spectral resolution for detection ($R_{\mathrm{Vis}} \ge 135$). These constraints are consistent with findings from past works \citep{des2002remote, brandt2014prospects, feng2018characterizing, damiano2022reflected}. While $R_{\mathrm{Vis}}$ should not drop below this recommendation, \autoref{tab:exposure} and Figures \ref{fig:exp_vs_res_sum} and \ref{fig:contour-combined} suggest that increased spectral resolution may slightly reduce the exposure time required to detect \ch{O2}, and we further discuss this implication in \autoref{sec:vis}.

In the near-IR, we are primarily concerned with detecting and/or constraining \ch{H2O}, \ch{CH4}, \ch{CO2}, and \ch{CO}. While our detectability calculation results (e.g., \autoref{tab:exposure}) show that $R_{\mathrm{NIR}}\ge20$ may be sufficient for detecting \ch{H2O} and \ch{CH4}, \ch{CO2} requires higher spectral resolving power ($R_{\mathrm{NIR}}\ge38$ in the early Archean, and $R_{\mathrm{NIR}}\ge66$ in the Phanerozoic). Though \ch{CO} seems to similarly require higher spectral resolutions for detection, we note that the minimum exposure time is uniformly very high across all Earth through time cases ($t_{\mathrm{exp}} \ge 10^6$ hours). For these cases, we lean more on the interpretation of our retrieval simulations, where information on upper limits is accessible. Our findings are consistent with those from previous works \citep{des2002remote, damiano2022reflected}. \autoref{fig:resolution_retrievals}d and \autoref{fig:covar} show that spectral resolutions $R_{\mathrm{NIR}}<40$ may introduce a concerning degeneracy between \ch{CO2} and \ch{CO}. This degeneracy results in the loss of a confident \ch{CO2} detection, important for characterizing planetary habitability, and a worsening upper limit constraints on \ch{CO}, an important antibiosignature for methanogenic life. We further discuss these results in \autoref{sec:nir}. A detailed comparison of our results to previous works is included in \autoref{sec:previous}. Finally, we discuss future research directions in \autoref{sec:future}.


\subsection{The Benefits of High Resolution in the Visible Require Lower Detector Noise}

\label{sec:vis}

At visible wavelengths, our detectability calculations and retrieval simulations suggest that the nominal resolution of $R_{\mathrm{Vis}} = 140$ is sufficient for detecting and characterizing \ch{O2} for the Phanerozoic Earth. Though the detectability calculations suggest that \ch{H2O} and \ch{O3} are generally over-resolved, this does not appreciably degrade our ability to constrain the abundance of these gases. Likewise, in our Phanerozoic retrieval simulations where \ch{O2}, \ch{O3}, and \ch{H2O} are confidently detected across all resolution cases, higher spectral resolution does not significantly enhance the precision on abundance constraints -- all $1\sigma$ uncertainties fall within $\pm$0.01 dex. {Similarly, cloud-top pressure shows minor gains in precision with increasing resolution.}

For the low-\ch{O2} Proterozoic atmospheres, our detectability calculations suggest that a minimum resolution of $R_{\mathrm{Vis}} = 171$ is required to detect \ch{O2} for exposure times within 20\% of the minimum, suggesting that the nominal $R_{\mathrm{Vis}} = 140$ is suboptimal for detecting Proterozoic-like \ch{O2} abundances. Accordingly, the nominal dark current and nominal resolution yield a minimum required exposure time of ${\sim}3000$ hours for detecting \ch{O2} at 0.76 \microns for such low abundances. To reduce the minimum required exposure time to detect \ch{O2} to 1000 hours, our contour plots (\autoref{fig:contour-combined}a suggest that the dark current would need to be reduced by at least an order of magnitude from the nominal HWO value ($3 \times10^{-5}$ to $1\times10^{-6}$), and the spectral resolution increased to nearly $R_{\mathrm{Vis}}=5000$. For more realistic exposure times that correspond to searches for modern-Earth \ch{O2} levels ($t_{\mathrm{exp}}{\sim}80$ hours; SNR$=20$ at 0.76 \microns for $R_{\mathrm{Vis}}=140$), our Proterozoic retrieval simulations suggest that higher spectral resolution does not meaningfully improve the precision of upper limits obtained on \ch{O2} or \ch{O3}. 

The detectors under consideration for HWO are already projected to improve upon Roman CGI by nearly two orders of magnitude \citep{stark2025cross}. Thus, the dark current we determined is required to detect \ch{O2} in low-\ch{O2} Proterozoic-Earth-like atmospheres at higher resolutions ($\sim$3 order of magnitude improvement over CGI) may ultimately be challenging with anticipated detector technology. Furthermore, even if such detectors are developed, the minimum required exposure time would only be reduced by at most a factor of $3$. Assuming comparable exposure times required for characterizing the Phanerozoic Earth, increasing the spectral resolution at the nominal dark current does not meaningfully improve our characterization of the atmosphere. For example, at the nominal resolution, the $3\sigma$ upper limit corresponds to an \ch{O2} abundance that is a factor of 9.3 larger than the true abundance. At the higher resolutions considered in our study, the $3\sigma$ upper limit corresponds to an \ch{O2} abundance that is a factor of 8.5 larger than the true abundance. In either case, the retrieved upper limits would not indicate whether \ch{O2} is present in this atmosphere. 

It is also important to recognize that spectral resolution decisions do not occur in a vacuum: resolution affects the detectability of all gases simultaneously, not just the target species of interest. For example, were $R_{\mathrm{Vis}}$ increased to $\geq$1000 to improve sensitivity to low \ch{O2} abundances, the exposure time required to detect \ch{H2O} would also increase by up to a factor of ${\sim}2$ at the nominal dark current (\autoref{fig:contour-combined}a). This is a meaningful penalty, given that \ch{H2O} is expected to be a near-ubiquitous feature of habitable-zone planets and would be a primary observational target across a full planetary survey. \citet{young2024retrievals} outline a decision-tree framework in which \ch{H2O} detection serves as the essential first step for establishing planetary habitability, with all subsequent observations, including searches for \ch{O2}, \ch{CH4}, and \ch{CO2}, contingent on that initial detection. Any potential gains for low-\ch{O2} worlds would therefore come at the cost of substantially more expensive observations for the more general case of planets with atmospheric water vapor. On the other hand, were $R_{\mathrm{Vis}}$ decreased to $\sim$30 to optimize for detecting \ch{H2O}, \ch{O2} would be under-resolved and our ability to accurately identify worlds with oxygenic life would be seriously compromised. Thus, optimizing spectral resolution for characterizing one gas may degrade the efficiency of characterizing others, and resolution requirements must be balanced across multiple species to inform instrument design. 

We conclude that the gains from higher spectral resolution in the visible are marginal at best, and that the nominal resolution in the visible ($R_{\mathrm{Vis}}=140$) is likely adequate for characterizing \ch{O2} on the the Earth through time, despite challenges associated with probing \ch{O2} in Proterozoic-Earth-like atmospheres. Furthermore, our analysis implies that such low-\ch{O2} atmospheres ($\leq$ 1\% PAL) would require more stringent detector noise properties to achieve the theoretical gains from high resolution, which could complicate technology risk assessments. Instead, potentially the most efficient path is to indirectly infer the presence of \ch{O2} via \ch{O3}, which may be readily detectable in the ultraviolet ($0.25 < \lambda <0.40$ \microns) at low spectral resolution ($R_{\mathrm{UV}}{\sim}7$) across a range of abundances \citep{schwieterman2018importance, gilbert2024retrieved, krissansen2025wavelength}. {By analogy, it is common with JWST NIRspec to observe at high resolution ($R{\sim}$1000) and bin observations down to lower resolution ($R{\sim}100$) for analysis \citep[e.g.,][]{carter2024benchmark}. Our results demonstrate this approach is only appropriate if the detector noise is sufficiently low. If the converse is true -- as is likely the case for Earth-like atmospheres in reflected light -- then the marginal information gain from higher resolution may not justify the additional noise incurred.}

\subsection{The Resolution in the Near-IR Should be Greater than 20 to Avoid Retrieval Ambiguities and Enable \ch{CO2} Detections}
\label{sec:nir}


In the near-IR wavelengths, our detectability calculation results suggest that spectral resolutions of $R_{\mathrm{NIR}}\geq42$ are required to detect \ch{CO2} for the early-through-late Archean Earth within our 20\% exposure time threshold. Given this exposure time threshold, we find that slightly higher spectral resolutions of $R_{\mathrm{NIR}}\geq61$ and $R_{\mathrm{NIR}}\geq66$ are better for detecting \ch{CO2} for the Proterozoic and Phanerozoic Earth, respectively. This suggests that the nominal $R_{\mathrm{NIR}} = 70$ is sufficient for detecting Proterozoic and Phanerozoic-like \ch{CO2} abundances. 

Overall, our results suggest that degrading the spectral resolution in the near-IR has modest effects on the characterization of Archean-like atmospheres. Our fixed-exposure-time retrieval simulations (\autoref{fig:resolution_retrievals}c), where \ch{CO2}, \ch{CH4}, and \ch{H2O} are confidently detected across all resolution cases, show that degrading the spectral resolution to as low as $R_{\mathrm{NIR}}=20$ does not significantly affect any of the abundance constraints on these gases -- our $1\sigma$ uncertainties never exceed $\pm0.01$ dex. Thus, though the detectability calculations suggest that \ch{H2O} and \ch{CH4} are generally over-resolved, this does not negatively impact our ability to constrain the abundance of these gases. Interestingly, the $3\sigma$ \ch{CO} upper limit generally increases with decreased spectral resolution. The $3\sigma$ upper limit of the coarsest resolution considered ($R_{\mathrm{NIR}} = 20$) is 0.19 dex larger than that of the nominal resolution case ($R_{\mathrm{NIR}} = 70$). These correspond to upper limits of 8.1 $\times 10^{3}$ ppm for the lowest resolution, and 1.6 $\times 10^{3}$ ppm for the nominal resolution. This trend for this particular Earth-through-time case suggests that lower spectral resolution may induce larger uncertainties on the abundance of undetected gases. {Cloud opacity and cloud fraction posteriors show similarly negligible losses in precision across all resolutions, even for the late Archean cloud fraction which shows the largest degradation relative to the nominal case.}

Our Phanerozoic retrieval simulations most clearly demonstrate how degrading the spectral resolution in the near-IR may lead to larger uncertainties on gas abundances. While decreasing the spectral resolution does not significantly impact our constraint on \ch{H2O} (within $\pm$0.01 dex) nor our upper limit on \ch{CH4} (within $\pm$0.03 dex), it does ultimately affect our ability to confidently detect \ch{CO2} and place upper limits on \ch{CO} abundances. As spectral resolution decreases, the $1\sigma$ uncertainties in the \ch{CO2} abundance constraints generally increase from $\pm0.14$ ($R_{\mathrm{NIR}}=70$) to $\pm0.20$ dex ($R_{\mathrm{NIR}}=40$). At the lowest resolution we investigated, $R_{\mathrm{NIR}}=20$, this increasing uncertainty ultimately gives rise to a concerning degeneracy between \ch{CO2} and \ch{CO} abundances. At this coarse spectral resolution, the \ch{CO2} detection degrades to an upper limit, and the 3$\sigma$ upper limit on \ch{CO} increases by 0.55 dex relative to the nominal case.  This effect occurs due to spectral degeneracies between \ch{CO2} and \ch{CO} absorption and a resulting covariance between their abundances (\autoref{fig:covar}). The lack of a \ch{CO2} detection in this case is concerning, as it is an important habitability indicator \citep{catlingdavid2018exoplanet}. Furthermore, the poorer upper limit on \ch{CO} results in worsening constraints on the ratio of \ch{CO}/\ch{CH4}. We note that the 3$\sigma$ upper limits on both gases for the lowest spectral resolution case do not allow us to rule out scenarios where \ch{CO}$>$\ch{CH4}, an important biosignature false positive diagnostic \citep{ragsdale2004life, zahnle2008photochemical, zahnle2011there, gao2015stability, wang2016detection, wogan2020chemical, krissansen2022understanding, ranjan2023importance}. {Cloud parameters are largely unaffected by resolution in this case; the Phanerozoic cloud opacity posterior degrades only slightly at the lowest resolution, and cloud fraction precision remains flat across all resolutions.}

We conclude that the nominal resolution in the near-IR ($R_{\mathrm{NIR}}=70$) is more than adequate for characterizing \ch{CH4}, \ch{CO2}, and \ch{H2O} on the Earth through time, and that spectral resolutions as low as $R_{\mathrm{NIR}}=40$ may be sufficient in this wavelength range. Our results also suggest that degrading the resolution in the near-IR below $R_{\mathrm{NIR}}=40$ may incur additional uncertainties in the abundance of \ch{CO2} and \ch{CO}. At lower spectral resolutions ($R_{\mathrm{NIR}}=20$), these additional uncertainties result in an ambiguity between low abundances of \ch{CO2} and high abundances of \ch{CO} which misrepresent the planetary atmosphere. We therefore conclude that resolutions $R_{\mathrm{NIR}}<40$ are demonstrably disadvantageous for accurately characterizing Earth-through-time atmospheres. {We note that the lowest resolution case also introduces a modest bias in the retrieved cloud opacity that is absent at higher resolutions; while minor in isolation, such biases could plausibly grow at even coarser resolutions or for atmospheric cases with more complex cloud structures, suggesting another possible disadvantage of operating below $R_{\mathrm{NIR}}=40$.}

\subsection{Comparison to Previous Work}
\label{sec:previous}

Our study assesses the spectral resolution needed to characterize Earth-through-time atmospheres using the same cases considered by \citet{krissansen2025wavelength}, with a long-wavelength cutoff of 1.7 \microns. Our retrieval results are broadly consistent despite key differences in study design. While \citet{krissansen2025wavelength} assumed constant SNR across the full wavelength range, we assumed constant exposure times. Additionally, we excluded UV wavelengths that they included.

For the Proterozoic Earth, \citet{krissansen2025wavelength} detected \ch{CO2} and \ch{CH4} with SNR ${>}10$ whereas our visible-focused retrievals obtained only an upper limit on both gases, likely because our $t_{\mathrm{exp}}{\sim}80$ hour simulations achieve SNR $<5$ in the near-IR, as shown in \autoref{fig:SNR_lambda}b. Both studies obtain upper limits on \ch{CO}, but neither detects \ch{O2}. We do not detect \ch{O3} due to our exclusion of UV wavelengths where \ch{O3} has strong absorption features. For the Archean Earth in the near-IR, both studies find that \ch{CO2} and \ch{CH4} are well-constrained. For the Phanerozoic Earth, both studies obtain upper limits on \ch{CH4} and constrain \ch{O2} and \ch{O3}. Finally, our \ch{CO2} results are consistent with \citet{krissansen2025wavelength} when accounting for SNR: in our visible-focused retrieval with low SNR in the near-IR, we obtain only upper limits as they do. In our near-IR-focused retrieval with SNR $>40$, we confidently detect \ch{CO2}, likely because our near-IR SNR exceeded the highest SNR case considered in their study.

In terms of spectral resolution, previous work argued that resolution in the visible should be set by the detection of the narrow-band \ch{O2} feature at 0.76 \microns \citep{des2002remote, brandt2014prospects, feng2018characterizing, luvoir2019luvoir, gaudi2020habitable, damiano2022reflected}, while the resolution in the near-IR should be set by the detection of \ch{CH4} and \ch{CO2} features \citep{des2002remote, luvoir2019luvoir, gaudi2020habitable, damiano2022reflected}. Here we compare our results to that of previous studies. Whereas our retrieval simulations are most comparable to those of \citet{feng2018characterizing} and \citet{damiano2022reflected}, our analytical calculations are more similar to those in \citet{brandt2014prospects}.  
Our retrieval simulations suggest that the nominal spectral resolution in the visible ($R_{\mathrm{Vis}}=140$) is sufficient to meaningfully constrain \ch{O2} for atmospheres akin to the Phanerozoic Earth. While our analytical calculations suggest that spectral resolutions up to 1,300 would also be acceptable for the nominal dark current, our retrieval results suggest that the information gained from higher resolutions is negligible compared to lower resolutions, given identical exposure times. Moreover, \autoref{fig:contour-combined} suggests that the required dark current needs to be at least an order of magnitude smaller for higher resolution to ensure equivalent oxygen detectability, further increasing the burden of detector complexity. These results are in agreement with \citet{feng2018characterizing} and \citet{damiano2022reflected}, who found that $R_{\mathrm{Vis}}=140$ (with SNR $=20$) is required in the optical wavelengths to ensure meaningful constraints on \ch{O2} abundance. Our results are also in agreement with the detectability calculations of \citet{brandt2014prospects}, which found that a resolution of 150 is required to detect \ch{O2}. By contrast, \citet{des2002remote} found that the required resolution to detect \ch{O2} in the optical and the near-IR is 69--72 by measuring the location and width of various atmospheric absorption features. The figure found by \citet{feng2018characterizing, damiano2022reflected}, and this work is likely twice as large due to the consideration of Nyquist sampling. 

In the near-IR, we found that a spectral resolution of $R_{\mathrm{NIR}}=40$ is sufficient for characterizing the Archean and Phanerozoic Earth atmospheres, given adequately long exposure times. However, this likely represents a lower limit on the spectral resolution in this wavelength range, as we found that $R_{\mathrm{NIR}}=20$ leads to ambiguities in the retrieved \ch{CO} and \ch{CO2} abundances. The LUVOIR final report \citep{luvoir2019luvoir} recommended $R_{\mathrm{NIR}}=70$ for \ch{CO2} detection with an integral field spectrograph and $R_{\mathrm{NIR}}=200$ for single-point spectrographs. Our detectability calculations for the Phanerozoic Earth are consistent with these recommendations, finding that a range of resolutions ($R_{\mathrm{NIR}}=66$--$564$) would yield \ch{CO2} detection in comparable exposure times ($t_{\mathrm{exp}}\approx3248$ hours). Rather than conducting retrieval simulations across this entire range, we focused on exploring the lower resolution limit suggested by our Archean Earth calculations ($R_{\mathrm{NIR}}=38$), finding that resolutions below $R_{\mathrm{NIR}}=40$ introduce degeneracies in atmospheric characterization. These results are consistent with those of \citet{damiano2022reflected}, who showed that a resolution of $R_{\mathrm{NIR}}=40$ in the near-IR is sufficient to characterize Archean/Modern Earths and discriminate \ch{CO2}-dominated atmospheres. Though our results are broadly consistent with the range of acceptable resolutions found by \citet{des2002remote}, $R_{\mathrm{NIR}}=11$--$40$, ($R_{\mathrm{NIR}}=22$--$80$ with Nyquist sampling) we rule out $R_{\mathrm{NIR}}<40$ for accurately characterizing the Phanerozoic Earth. \citet{brandt2014prospects} and \citet{feng2018characterizing} did not assess spectral resolution in the near-IR, so we are not able to compare our results to these prior works.

\subsection{Future Work}
\label{sec:future}

In this study, we explored the effects of assumed spectral resolution in the visible and near-IR wavelengths for a limited set of atmospheric cases representing the Earth through time. We did not explore the effect of spectral resolution in the UV, and we did not consider a broader range of atmospheric cases, which include abiotic false positive atmospheres. We also assumed photon-limited observations throughout, such that our absolute exposure time estimates do not account for time-independent noise sources like speckle noise that could introduce a noise floor at long integration times; our predictions of relative changes in exposure time across resolution cases are therefore more robust than the absolute values. Finally, we did not investigate the sensitivity of spectral resolution for surface characterization, including surface biosignatures, nor the effects of wavelength-dependent clouds with variable cloud coverage. Future work should address these gaps to ensure HWO spectroscopy is fully capable of identifying living planets and discriminating their imitators. 

Future work should explore the spectral resolution in the UV needed to detect and constrain \ch{O3} abundances. As a photochemical byproduct of \ch{O2}, \ch{O3} may indirectly indicate the presence of \ch{O2} in low-\ch{O2} atmospheres like the Proterozoic Earth \citep{krissansen2025wavelength}. Though it has weak absorption features in the visible wavelengths, the Hartley-Huggins bands in the UV are most sensitive to changes in abundance and the presence of a stratospheric bulge \citep{gilbert2024retrieved}. Given our results showing that higher spectral resolution in the visible cannot improve upper limits on \ch{O2} abundance, UV capabilities may be particularly important for characterizing Proterozoic-like, low-\ch{O2} atmospheres \citep{damiano2023reflected, young26, washington26}. 

Our study did not assess spectral resolution for a broad range of atmospheric cases and species despite the potential diversity of terrestrial exoplanet atmospheres to be discovered. Future work should extend this assessment to additional atmospheric cases, including false positive atmospheres. For example, recent work has shown that false positive atmospheres can inform HWO instrument requirements like the short-wavelength cutoff in the UV and the long-wavelength cutoff in the near-IR \citep{krissansen2025wavelength}. Similarly, future work should assess how spectral resolution affects our ability to constrain \ch{SO2}, an intermediate product of sulfate metabolisms with absorption features that overlap with those of \ch{O3} in the near-UV \citep{young26, washington26}. Another species that should be considered in future works is \ch{N2O}, a waste product of microbial nitrogen metabolisms with absorption features that overlap with other near-IR absorbers \citep{tokadjian2024detectability}. Broadening the {types of} atmospheres {(e.g., temperate/cold sub-Neptunes, temperate/cold gas giants)} and species under investigation is particularly important for a complete assessment of how spectral resolution affects our sensitivity to spectral degeneracies.  


Our exposure time calculations assumed photon-limited observations, but real observations may deviate from this regime in ways that affect the absolute exposure times reported here, though the relative changes in exposure time across resolution cases are likely more robust. Time-independent noise sources, such as speckle noise from residual starlight, introduce a noise floor that photon-limited scaling cannot capture. At sufficiently long exposure times, such a noise floor could prevent detection of species that already require exceedingly long integration times ($>$100 hours) at the nominal resolutions. This caveat is particularly relevant to our Proterozoic results, where the exposure times required to detect \ch{O2} ($t_{\mathrm{exp}}{\sim}2000$ hours) already exceed practical limits. In this context, our predictions of \textit{relative} changes in exposure time across resolution cases are more robust than the absolute exposure times themselves, which will also vary with target distance and telescope aperture. Future work should incorporate time-independent noise components to assess whether spectral resolution choices remain optimal in the presence of a noise floor, and to refine exposure time estimates for specific HWO target stars. 

Speckles can also generate correlated noise across the detector as a function of wavelength, which can mimic or obscure absorption features. While we do not account for correlated noise in this study, see \citet{ruffio2026characterizing} for a detailed investigation. Despite neglecting speckles, our findings agree with \citet{ruffio2026characterizing} in that sufficiently low detector noise can enable more sensitive, high resolution ($R{\sim}1000$) observations, particularly for narrow-band absorption features. \citet{ruffio2026characterizing} highlight that ultra-stable wavefront control minimizes correlated speckle noise, but this level of performance poses a significant technological challenge for HWO. Future work could explore the trade space between relaxed (stricter) wavefront stability requirements and stricter (relaxed) detector noise requirements needed for high (low) resolution observations \citep[e.g., using][]{steiger2026incorporating}.

Our investigation also assumed gray surface albedos throughout, but recent work has demonstrated that reflected light observations can be sensitive to surface composition \citep{ulses2025detecting}, including biological pigments \citep{hegde2015surface, schwieterman2018exoplanet, gomez2023search, borges2024detectability}. \citet{gomez2023search} showed that wavelength-dependent retrieval models can accurately identify spectral edge features, including the ``vegetative red edge'' -- a sharp albedo transition near 750 nm caused by vegetation. Constraining the surface composition of a planet may also help discriminate abiotic scenarios. For example, \citet{ulses2025detecting} demonstrated that reflected light observations spanning 0.3--1.0 \microns can reveal the presence of land surfaces, which would help rule out water-world false positives. Future work should investigate how spectral resolution affects the detectability of wavelength-dependent albedo features and determine the UV, visible, and near-IR resolutions needed to break potential spectral degeneracies, such as the \ch{O3}-land degeneracy identified by \citet{ulses2025detecting}.

{Finally, our retrieval simulations also adopted simplifying assumptions about clouds: we assumed a constant cloud fraction across all atmospheric cases and treated clouds as gray absorbers. While these choices provide a useful controlled baseline, real cloud populations exhibit wavelength-dependent absorption features that can be degenerate with gas-phase absorbers. For example, \citet{gilbert2024retrieved} demonstrated that ice cloud absorption features in the near-IR can produce false positive detections of gases such as \ch{CO}. In principle, higher spectral resolution may help break such degeneracies, since the sharp pressure-broadened line structure of gas-phase species is readily distinguishable from the broad, featureless absorption of condensed-phase particles at sufficient resolution. Future work should investigate how varying cloud fraction, cloud particle composition, and particle size affect retrieval outcomes, and assess whether the resolution requirements identified here remain robust across a broader range of cloud scenarios.}

\section{Conclusion}
\label{sec:conclusion}

We assessed how spectral resolution affects HWO's ability to detect and characterize biosignature gases in Earth-like exoplanet atmospheres using complementary detectability calculations and atmospheric retrievals. Our analysis focused on \ch{O2}, \ch{O3}, \ch{H2O}, \ch{CH4}, \ch{CO2}, and \ch{CO} across visible and near-IR wavelengths for atmospheres representing Earth's evolution from the Archean through the Phanerozoic.

In the visible, we find that the nominal resolution of $R_{\mathrm{Vis}}=140$ is sufficient for detecting \ch{O2} in Phanerozoic-like atmospheres. While higher resolutions could theoretically reduce the minimum exposure time needed to detect \ch{O2} in low-\ch{O2} Proterozoic atmospheres, the required improvements in detector noise (${>}10\times$ reduction in dark current from our nominal value assumed for HWO) may increase technological risk. Furthermore, spectral resolution decisions do not occur in a vacuum: increasing the spectral resolution in the visible to improve sensitivity for low-\ch{O2} worlds could increase the exposure time required to detect \ch{H2O} by up to a factor of ${\sim}2$, penalizing the foundational habitability constraint that anchors downstream biosignature search strategies. The marginal improvements in exposure time seen for \ch{O2} by increasing $R_{\mathrm{Vis}}$  likely do not justify significant increases in spectral resolution beyond the nominal value, although small increases beyond 140 have no major downsides. In the near-IR, we find that $R_{\mathrm{NIR}}\geq40$ is necessary to avoid a concerning degeneracy between \ch{CO2} and \ch{CO} that could produce false positive detections of abundant \ch{CO}. The nominal resolution of $R_{\mathrm{NIR}}=70$ provides robust constraints on \ch{CO2} and \ch{H2O} across all Earth-through-time cases.

These results support HWO's current baseline resolution choices. Future work should extend this analysis to UV wavelengths to assess \ch{O3} detection in low-\ch{O2} atmospheres and to a broader range of atmospheric scenarios including abiotic false positive cases. Our findings provide actionable guidance, including some flexibility, for finalizing HWO's spectrometer requirements while maintaining the technical capability necessary for the search for life on exoplanets.

\begin{acknowledgments}
    This work was inspired by and benefited from conversations within the HWO Living Worlds working group. We thank Christopher Stark, Natasha Latouf, and Jean-Baptiste Ruffio for informative discussions at the Towards the Habitable Worlds Observatory symposium in the summer of 2025, which helped guide this investigation. We also acknowledge productive discussions with Natasha Batalha, Nicholas Wogan, Brianna Lacy, and Kevin Zahnle. We thank Tyler Robinson and one anonymous reviewer for their helpful comments, which improved this manuscript. We also thank Stephanie Olson for the use of her dedicated nodes on the Bell cluster at the Purdue University Rosen Center for Advanced Community (RCAC), as well as RCAC's IT personnel \citep{McCartney2014}. This work was supported by NASA Astrophysics Decadal Survey Precursor Science grant 80NSSC23K1471. JKT was supported by the NASA Nexus for Exoplanet System Science (NExSS), funded via the NASA Astrobiology Program grant No. 80NSSC23K1398, and the Alfred P. Sloan Foundation under grant No. 2025-25204. JLY acknowledges internal support from Johns Hopkins APL. We used Claude (Anthropic) to assist with \LaTeX\ table formatting and finalizing plotting scripts.
\end{acknowledgments}

\software{\rfast \citep{robinson2023exploring}, \pyedith \citep{Alei2025}, \emcee \citep{foreman2013emcee}, Claude \citep{claude2025}} 

\bibliography{bib}{}
\bibliographystyle{aasjournal}

\appendix

\vspace{-20pt}

\begin{figure}[H]
    \includegraphics[width=0.9\linewidth]{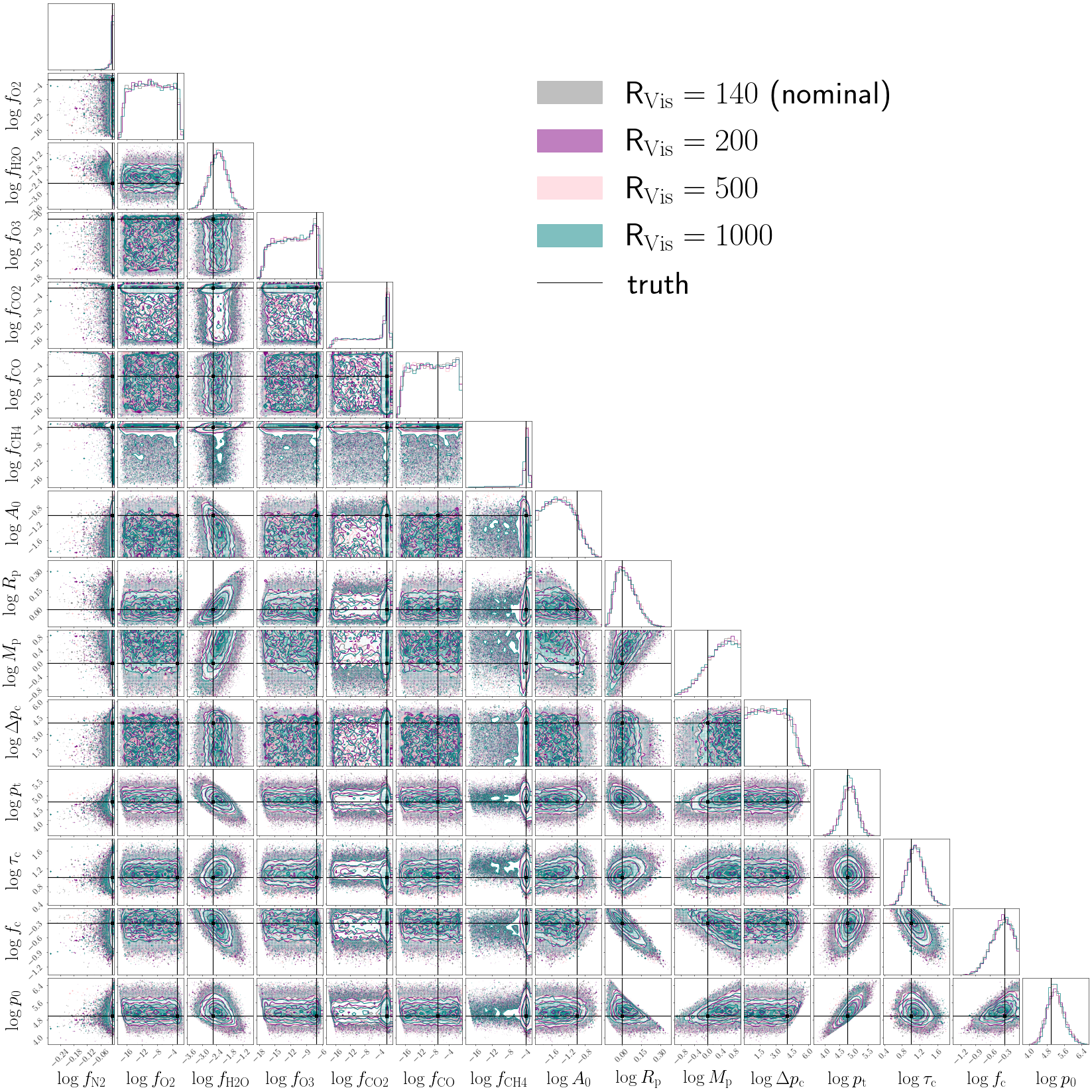}
    \caption{Full corner plot for the high-\ch{CH4} Proterozoic case, including the nominal spectral resolution ($R_{\mathrm{Vis}}=140$) and higher resolutions ($R_{\mathrm{Vis}}=200$, 500, and 1000) in the visible (VIS) wavelengths.}
    \label{fig:corner1}
\end{figure}

\begin{figure}[H]
    \includegraphics[width=0.9\linewidth]{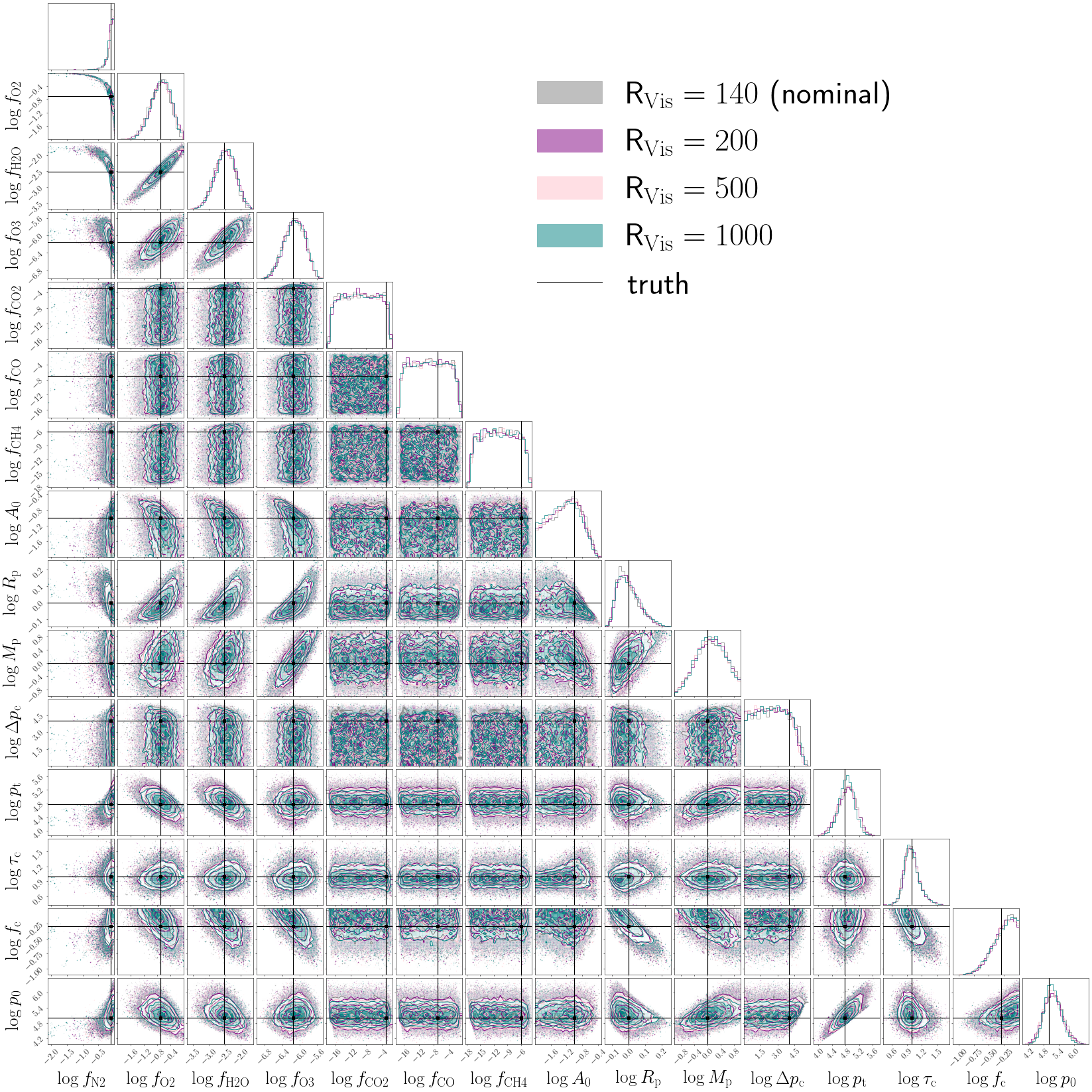}
    \caption{Full corner plot for the Phanerozoic case, including the nominal spectral resolution ($R_{\mathrm{Vis}}=140$) and higher resolutions ($R_{\mathrm{Vis}}=200$, 500, and 1000) in the visible (VIS) wavelengths.}
    \label{fig:corner2}
\end{figure}

\begin{figure}[H]
    \includegraphics[width=0.9\linewidth]{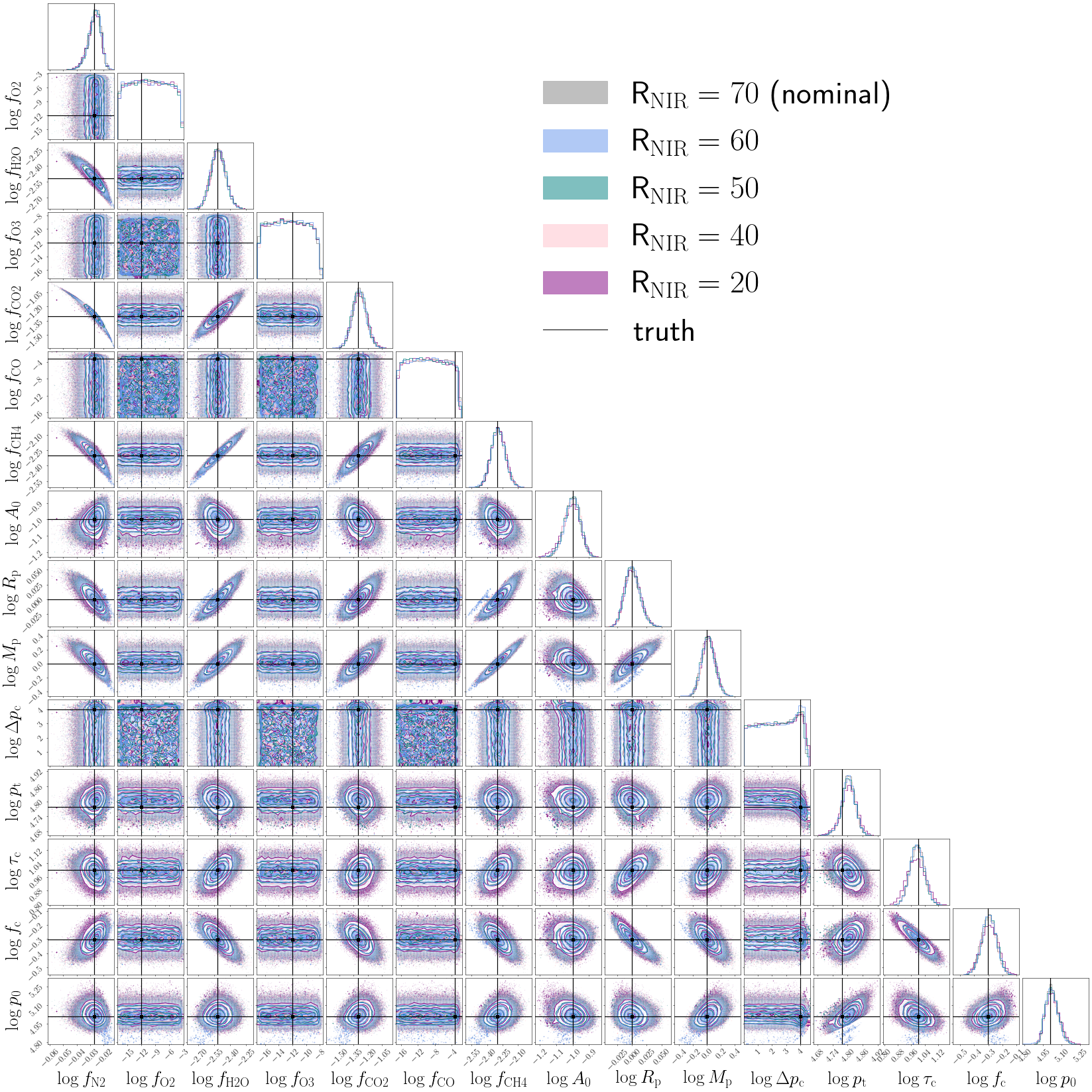}
    \caption{Full corner plot for the Late Archean case, including the nominal spectral resolution ($R_{\mathrm{NIR}}=70$) and lower resolutions ($R_{\mathrm{Vis}}=60$, 50, 40, and 20) in the near-infrared (NIR) wavelengths.}
    \label{fig:corner3}
\end{figure}

\begin{figure}[H]
    \includegraphics[width=0.9\linewidth]{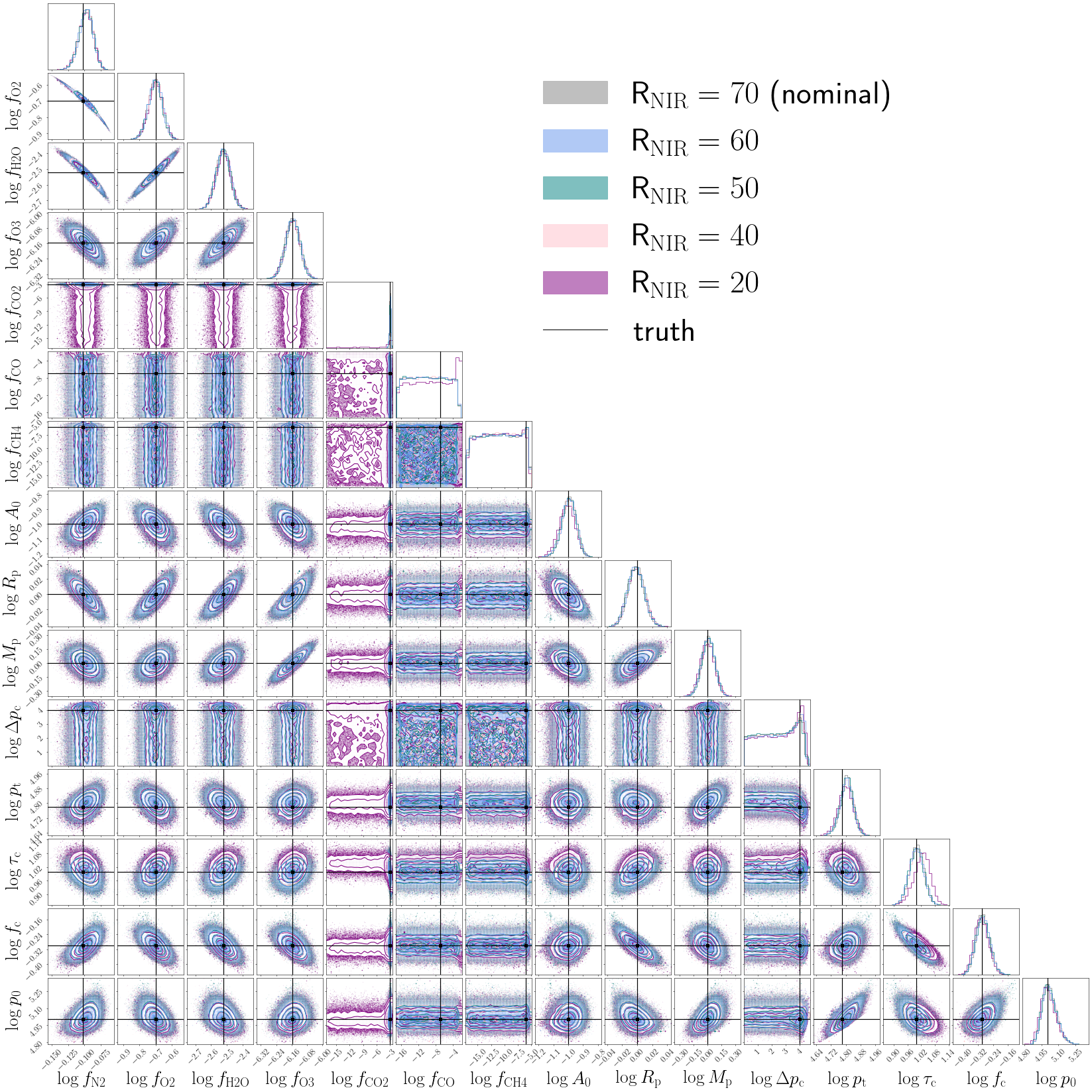}
    \caption{Full corner plot for the Phanerozoic case, including the nominal spectral resolution ($R_{\mathrm{NIR}}=70$) and lower resolutions ($R_{\mathrm{Vis}}=60$, 50, 40, and 20) in the near-infrared (NIR) wavelengths.}
    \label{fig:corner4}
\end{figure}

{In Figures \ref{fig:corner1}, \ref{fig:corner2}, \ref{fig:corner3}, and \ref{fig:corner4} we show the full corner plots for each atmospheric case explored in the retrieval portion of this investigation. Each plot overlays the different spectral resolution cases for a given Earth-through-time atmosphere, including the corresponding ``truth'' values for each retrieval parameter. In addition to the atmospheric abundances shown in the main text, the full corner plots include surface albedo ($A_0$), planet radius ($R_p$) and mass ($M_p$) in Earth units, cloud layer thickness ($\Delta p_c$) in Pa, cloud-top pressure ($p_t$) in Pa, cloud opacity ($\tau_c$), cloud fraction ($f_c$), and total atmospheric surface pressure ($p_0$) in Pa. Since all gases are retrieved originally in partial pressure space, the total surface pressure posterior represents the sum-total of the gas posteriors. A brief discussion of key cloud parameters is included in the main text, with a more thorough discussion below.}

\newpage

\begin{figure}[H]
\centering
    \includegraphics[width=\linewidth]{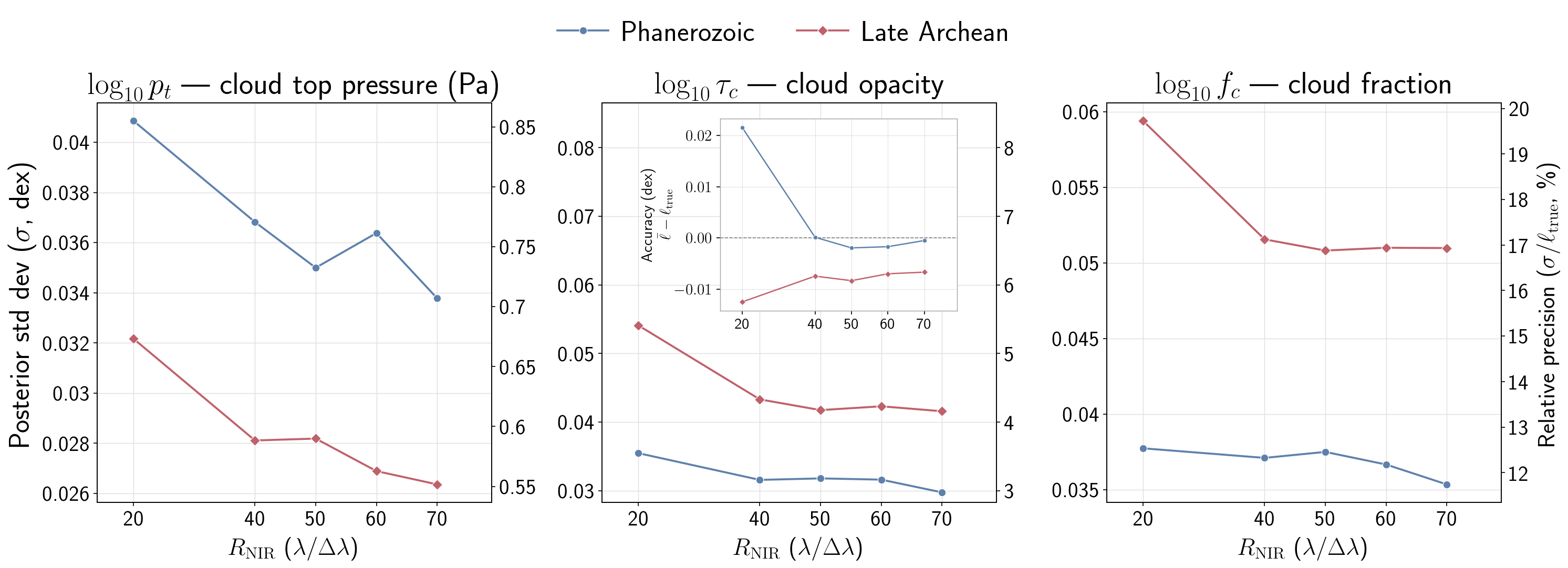}
    \caption{Cloud parameter retrieval diagnostics for the near-IR channel as a function of spectral resolution for the Phanerozoic (blue circles) and Late Archean (red diamonds) atmospheric cases. Each column shows a different cloud parameter: cloud-top pressure ($p_t$), cloud opacity ($\tau_c$), and cloud fraction ($f_c$). The left-hand y-axis shows the posterior standard deviation and the right-hand y-axis shows the relative precision ($\sigma/\ell_\mathrm{true}$) of the retrieved parameter, both computed in log$_{10}$ space. The inset in the $\tau_c$ standard deviation panel shows the retrieval accuracy ($\bar{\ell} - \ell_\mathrm{true}$) for cloud opacity, highlighting the minor systematic bias that emerges at the lowest resolution ($R_{\mathrm{NIR}}=20$) for the Phanerozoic case. The nominal resolution ($R_{\mathrm{NIR}}=70$) is marked on the x-axis of each panel for reference.}
    \label{fig:cloud_nir}
\end{figure}

\begin{figure}[H]
\centering
    \includegraphics[width=0.6\linewidth]{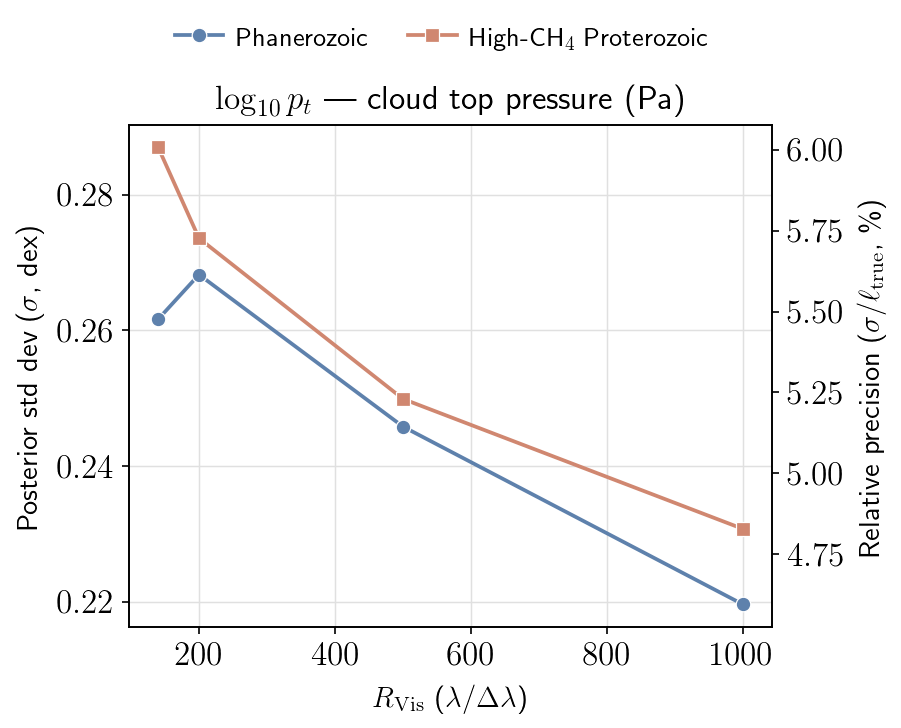}
    \caption{Cloud-top pressure ($p_t$) retrieval diagnostics for the visible channel as a function of spectral resolution for the Phanerozoic (blue circles) and High-CH$_4$ Proterozoic (orange squares) atmospheric cases. The left-hand y-axis shows posterior standard deviation and the right-hand y-axis shows relative precision ($\sigma/\ell_\mathrm{true}$), both computed in log$_{10}$ space. Both atmospheric cases show modest but consistent improvements in retrieval precision with increasing spectral resolution.}
    \label{fig:cloud_vis}
\end{figure}

{Figures \ref{fig:cloud_nir} and \ref{fig:cloud_vis} highlight minor trends in the retrieved cloud parameters, which are similar to those detailed for the gas abundance parameters in the main text. We show the retrieval accuracy (median $\bar{\ell}$ - truth ${\ell_{\mathrm{true}}}$), posterior standard deviation ($\sigma$), and the relative precision ($\sigma$/${\ell_{\mathrm{true}}}$) for select cloud parameters. Across all the Earth-through-time cases in the visible and near-IR, we find $\sim$0.03--0.08 dex improvement in the standard deviation of the retrieved cloud top pressure posterior, corresponding to a 0.1--1\% improvement in relative precision, as a function of increased spectral resolution. For the Late Archean and Phanerozoic atmospheres explored in the near-IR, the standard deviation (\autoref{fig:cloud_nir}, left panel) of the retrieved cloud opacity posterior remains mostly flat across the spectral resolutions under investigation. However, for the lowest resolution case ($R_{\mathrm{NIR}}$=20) for the Phanerozoic and Archean atmospheres, we find that the posteriors are less precise than those of the nominal resolution ($R_{\mathrm{NIR}}$=70) case by 0.005 and 0.01 dex in standard deviation, corresponding to 0.5\% and 1.2\% worse relative precision, respectively.} 

\vspace{-40pt}

{The Phanerozoic cloud opacity also exhibits a larger systematic bias at the lowest resolution, with the posterior median deviating from the true value by 2.2\% at $R_{\mathrm{NIR}}$=20 compared to only 0.1\% at the nominal resolution, suggesting that coarse spectral resolution may introduce modest biases in cloud retrievals in addition to the precision losses described above (\autoref{fig:cloud_nir}, middle panel, inset). Finally, while \autoref{fig:cloud_nir} (right panel) shows that the standard deviation of the retrieved cloud fraction posterior remains flat with spectral resolution for the Phanerozoic atmosphere in the near-IR, the late Archean case shows a loss of precision of 0.08 dex in standard deviation corresponding to a loss of 3\% relative precision for the lowest resolution case relative to the nominal resolution. Although Figures \ref{fig:cloud_nir} and \ref{fig:cloud_vis} do reveal improvements in the retrieved precision of the cloud parameters at higher resolutions, these gains are sufficiently small so as to be nearly imperceptible in the corner plots (Figures \ref{fig:corner1}, \ref{fig:corner2}, \ref{fig:corner3}, and \ref{fig:corner4}) where these posteriors are over-plotted. Therefore, it is not evident that such gains would lead to statistically significant improvements in practice.}

\end{document}